\documentclass[useAMS,usenatbib,usegraphicx]{mn2e}
\input {psfig.sty}
\newcommand{\etal}{{et al.~}}

\newcommand{\kms}{\>{\rm km}\,{\rm s}^{-1}}

\newcommand{\Mpc}{\>{\rm Mpc}}

\newcommand{\mpch}{\>h^{-1}{\rm {Mpc}}}

\newcommand{\Msunh}{\>h^{-1}\rm M_\odot}
\newcommand{\Lsunh}{\>h^{-2}\rm L_\odot}

\newcommand{\beq}{\begin{equation}}
\newcommand{\eeq}{\end{equation}}

\newcommand{\rmd}{{\rm d}}

\newcommand{\Lsunhh}{\,h^{-2}\rm L_\odot}

\newcommand{\lcen}{L_{\rm c}}
\newcommand{\lc}{L_{\rm c}}
\newcommand{\avg}[1]{\langle #1 \rangle}
\newcommand{\avglogm}{\avg{\log M}(\lcen)}
\newcommand{\avgloglc}{\avg{\log \lcen}(M)}
\newcommand{\siglogm}{\sigma_{\log M}(\lcen)}

\newcommand{\drm}{{\rm d}}
\newcommand{\pdv}{{P (\Delta V)}}
\newcommand{\dv}{{\Delta V}}
\newcommand{\sigcen}{\sigma_{\log L}}
\newcommand{\sigsw}{\sigma_{\rm sw}}
\newcommand{\sighw}{\sigma_{\rm hw}}
\newcommand{\sigsat}{\sigma_{\rm sat}}
\newcommand{\avgsigsatsq}{\avg{\sigsat^2}}
\newcommand{\avnsat}{\avg{N_{\rm sat}}}
\newcommand{\avnsatm}{\avnsat(M)}
\newcommand{\philcm}{\Phi_{\rm c}(L|M)}
\newcommand{\philsm}{\Phi_{\rm s}(L|M)}
\newcommand{\plcm}{P(\lcen|M)}
\newcommand{\pmlc}{P(M|\lcen)}

\def\gtsima{$\; \buildrel > \over \sim \;$}
\def\ltsima{$\; \buildrel < \over \sim \;$}
\def\prosima{$\; \buildrel \propto \over \sim \;$}
\def\gsim{\lower.7ex\hbox{\gtsima}}
\def\lsim{\lower.7ex\hbox{\ltsima}}
\def\simgt{\lower.7ex\hbox{\gtsima}}
\def\simlt{\lower.7ex\hbox{\ltsima}}
\def\simpr{\lower.7ex\hbox{\prosima}}

\newdimen\hssize
\newdimen\hdsize
\begin{document}
\setlength{\hbadness}{10000}
\setlength{\vbadness}{10000}
%%%%%%%%%%%%%%%%%%%%%%%%%%%%%%%%%%%%%%%%%%%%%%%%%%%%%%%%%%%%%%%%%%%%%%%%%%

\title[Satellite Kinematics II]
        {Satellite Kinematics II: The Halo Mass-Luminosity Relation
 of Central Galaxies in SDSS}
\author[More \etal]
       {Surhud More$^{1}$ \thanks{International Max Planck
Research School fellow \newline Email: more@mpia.de},
Frank C. van den Bosch$^{1}$, Marcello Cacciato$^{1}$, H. J. Mo$^{2}$,
\and Xiaohu Yang$^{3}$, Ran Li$^{2,4}$ \\
        $^{1}$Max Planck Institute for Astronomy, Koenigstuhl 17,
D69117, Heidelberg, Germany\\
        $^{2}$Department of Astronomy, University of Massachussets,
Amherst, MA 010039305, USA\\
        $^{3}$Shanghai Astronomical Observatory, Nandan Road 80,
Shanghai 200030, China\\
        $^{4}$Department of Astronomy, Peking University, Beijing 100871, China}

%%%%%%%%%%%%%%%%%%%%%%%%%%%%%%%%%%%%%%%%%%%%%%%%%%%%%%%%%%%%%%%%%%%%%%%%%%

\date{}

\maketitle

\label{firstpage}

%%%%%%%%%%%%%%%%%%%%%%%%%%%%%%%%%%%%%%%%%%%%%%%%%%%%%%%%%%%%%%%%%%%%%%%%%%

\begin{abstract}
  The  kinematics of  satellite  galaxies reflect  the  masses of  the
  extended dark matter haloes in which they orbit, and thus shed light
  on the mass-luminosity relation (MLR) of their corresponding central
  galaxies.  In  this paper we select  a large sample  of centrals and
  satellites from the Sloan Digital  Sky Survey (SDSS) and measure the
  kinematics  (velocity dispersions)  of the  satellite galaxies  as a
  function of the $r$-band  luminosity of the central galaxies.  Using
  the analytical framework presented in  Paper I, we use these data to
  infer {\it  both} the  mean and  the scatter of  the MLR  of central
  galaxies, carefully  taking account of selection  effects and biases
  introduced  by  the   stacking  procedure.   As  expected,  brighter
  centrals on average reside in  more massive haloes.  In addition, we
  find  that  the scatter  in  halo masses  for  centrals  of a  given
  luminosity,  $\sigma_{\log  M}$,   also  increases  with  increasing
  luminosity.  As we demonstrate,  this is consistent with  $\sigcen$,
  which reflects  the scatter in the  conditional probability function
  $\plcm$,  being  independent of  halo  mass.   Our  analysis of  the
  satellite  kinematics  yields  $\sigcen=0.16\pm0.04$,  in  excellent
  agreement with constraints from clustering and group catalogues, and
  with predictions  from a semi-analytical model of  galaxy formation.
  We  thus  conclude  that  the  amount  of  stochasticity  in  galaxy
  formation, which is characterized by $\sigcen$, is well constrained,
  is independent of  halo mass, and is in  good agreement with current
  models of galaxy formation.
\end{abstract}

%%%%%%%%%%%%%%%%%%%%%%%%%%%%%%%%%%%%%%%%%%%%%%%%%%%%%%%%%%%%%%%%%%%%%%%%%%

\begin{keywords}
galaxies: haloes ---
galaxies: kinematics and dynamics ---
galaxies: fundamental parameters ---
galaxies: structure ---
dark matter ---
methods: statistical 
\end{keywords}

%%%%%%%%%%%%%%%%%%%%%%%%%%%%%%%%%%%%%%%%%%%%%%%%%%%%%%%%%%%%%%%%%%%%%%%%%%

\section{Introduction}
\label{sec:intro}

According to  the standard framework of galaxy  formation, dark matter
haloes form  gravitational potential wells in  which baryons collapse,
dissipate    their    energy    and    form   stars    and    galaxies
\citep{White1978,Blumenthal1984}.   The  complex  process   of  galaxy
formation and evolution is believed to  be governed by the mass of the
dark  matter halo  in which  it occurs.  To understand  the  halo mass
dependence of  this process, it  is important to  statistically relate
the observable properties (e.g.  luminosity) of galaxies to the masses
of their dark matter haloes.

The dark matter halo is bigger  in extent than the visible matter in a
galaxy  due  to  the inability  of  dark  matter  to lose  energy  via
dissipative  processes.  Therefore,  a tracer  population, distributed
over  large distances  from the  center of  the halo  is  necessary to
reliably measure  the mass of  a dark matter halo.  Satellite galaxies
satisfy this requirement and their kinematics reflect the depth of the
potential  well they orbit.   However, a  reliable measurement  of the
kinematics requires a large number of satellites. Whereas a relatively
large number  of satellites can  be detected in  cluster-sized haloes,
only a handful  of satellites can be detected in  less massive haloes. 
This leads  to a  large uncertainty on  the kinematic  measurement for
less  massive  haloes.  However,  under  the  assumption that  central
galaxies of similar properties reside  in similar mass haloes, one can
stack  such central  galaxies and  use  their satellites  to obtain  a
statistical  measure  of the  kinematics  even  in  low mass  haloes.  
\citep{Erickson1987,Zaritsky1993,Zaritsky1994,Zaritsky1997}.

During the  past few  years, data from  large galaxy  redshift surveys
such as the Sloan Digital Sky Survey \citep[SDSS;][]{York2000} and the
Two degree Field Galaxy Redshift Survey \citep[2dFGRS;][]{Colless2001}
has become  available. The  kinematics of satellite  galaxies selected
from these large datasets have  been used to infer the mass-luminosity
relationship      (hereafter     MLR)     of      central     galaxies
\citep{McKay2002,Brainerd2003,    Prada2003,   van    den   Bosch2004,
  Conroy2005,Conroy2007}  and to  study the  density profiles  of dark
matter    haloes    \citep{Prada2003,Klypin2007}.     These    studies
consistently  find that  the  velocity dispersion  of satellites  (and
hence  the mass  of the  halo) increases  with the  luminosity  of the
central   galaxy.    However,   \citet[][hereafter   N08]{Norberg2008}
demonstrated  a  quantitative disagreement  in  the  results of  these
studies and  showed that  this disagreement is  largely due  to subtle
differences  in the selection  criteria used  to identify  central and
satellite  galaxies.   Therefore,  it  is crucial  to  understand  how
selection  effects bias  the  MLR of  central  galaxies inferred  from
satellite kinematics.

The MLR of  central galaxies is specified in  terms of the conditional
probability  $\pmlc$, which  describes the  probability for  a central
galaxy with  luminosity $\lc$ to reside in  a halo of mass  $M$. For a
completely  deterministic  relation  between  halo  mass  and  central
luminosity,   $\pmlc=\delta(M-M_0)$,   where   $\delta$  denotes   the
Dirac-delta  function   and  $M_0$  is  a   characteristic  halo  mass
corresponding   to  centrals   of  luminosity   $\lc$.   The  velocity
dispersion,   $\sigma(\lc)$,  measured   by  stacking   centrals  with
luminosity $\lc$, then  translates into a mass $M_0$  according to the
scaling relation $\sigma^3 \propto  M$. However, galaxy formation is a
stochastic process  and the distribution  $\pmlc$ is expected  to have
non-zero  scatter. If this  scatter is  appreciable then  the stacking
procedure  results in combining  the kinematics  of haloes  spanning a
wide range  in halo mass.  This complicates the interpretation  of the
velocity dispersion.  We addressed this  issue in More, van  den Bosch
\& Cacciato (2008; hereafter Paper I), where we investigated a method
to  measure both  the  mean and  the  scatter of  the  MLR of  central
galaxies  using  satellite   kinematics.  We  outlined  two  different
weighting schemes  to measure  the velocity dispersion  of satellites,
{\it  satellite-weighting} and {\it  host-weighting}, and  showed that
the mean and  the scatter of the MLR can be  inferred by modelling the
velocity dispersion measurements in  these two schemes simultaneously. 
In  this paper,  we  apply  this method  to  the spectroscopic  galaxy
catalogue from  SDSS (Data Release 4)  in order to  determine both the
mean and the scatter of the MLR of central galaxies.

The aim  of this  paper is twofold.  First, we  carry out a  series of
tests on a  realistic mock galaxy catalogue to  validate our method to
infer  the  MLR of  central  galaxies  from  a redshift  survey  using
satellite   kinematics.     In   particular,   we    show   that   our
central-satellite selection  criterion and  the method to  measure the
kinematics reliably  recovers the true kinematics present  in the mock
catalogue.  We  also show  that the  mean and the  scatter of  the MLR
inferred from the kinematics match the corresponding true relations in
the mock  catalogue. Second, we  repeat the analysis on  galaxies from
the SDSS. In particular, we show that both the mean and the scatter of
the  MLR increase  with the  luminosity of  the central  galaxy.  This
demonstrates  that the scatter  in halo  masses is  non-negligible and
must be  taken into  account when interpreting  measurements involving
stacking.

This paper  is organized  as follows. In  Section~\ref{sec:mockcat} we
describe the construction  of the mock catalogue that  is used to test
our  method  of  analysis.  In Section~\ref{sec:selncrit}  we  briefly
outline the iterative selection  criterion used to select centrals and
satellites.  In  Section~\ref{sec:veldisp} we  describe  and test  the
method used to  measure the kinematics of satellites  as a function of
the central luminosity.  The inference of the MLR  from the kinematics
of  satellite galaxies requires  the knowledge  of the  number density
distribution     of      satellites     within     a      halo.     In
Section~\ref{sec:nsatprof} we show that  this distribution can also be
inferred from  the selected satellites.  In Section~\ref{sec:model} we
describe our model to  interpret the measured velocity dispersions and
show that this  model is able to recover the true  mean and scatter of
the   MLR  of   central   galaxies  from   the   mock  catalogue.   In
Section~\ref{sec:sdssapp}  we apply  our analysis  method to  a volume
limited sample  from SDSS  and discuss the  results. We  summarize our
results in Section~\ref{sec:summary}.

\section{Mock catalogues}
\label{sec:mockcat}

It is  important to carefully identify central  and satellite galaxies
from a redshift  survey in order to study  the kinematics of satellite
galaxies. Furthermore,  it is also important to  reliably quantify the
kinematics  of  the  selected  satellites  as a  function  of  central
luminosity which  in turn  can yield the  MLR of central  galaxies. We
monitor the  performance of our method  of analysis for  each of these
tasks using a realistic mock  galaxy catalogue (MGC) which serves as a
control dataset. The halo occupation of galaxies in the MGC is known a
priori,  thereby  allowing an  accurate  assessment  of  the level  of
contamination of the selected sample of centrals and satellites due to
false  identifications and  also a  comparison between  the kinematics
recovered  from  the selected  satellites  and  the actual  kinematics
present in  the MGC. 

The  two  essential  steps  to   construct  a  MGC  are  to  obtain  a
distribution of dark matter haloes and to use a recipe to populate the
dark matter  haloes with  galaxies. For the  former purpose, we  use a
numerical simulation of dark matter particles in a cosmological setup.
For the latter, we use the conditional luminosity function (CLF) which
describes  the average  number of  galaxies with  luminosities  in the
range $L \pm \drm L/2$ that reside in a halo of mass $M$.

A  distribution of  dark matter  haloes  is obtained  from a  $N$-body
simulation for a  $\Lambda$CDM cosmology with parameters, $\Omega_{\rm
  m}=0.238, \Omega_\Lambda = 0.762, \sigma_8 = 0.75, n_{\rm s} = 0.95,
h =  H_0/100 \kms\Mpc^{-1}  = 0.73$. The  simulation consists of  $N =
512^3$ particles  within a cube  of side $L_{\rm box}=300  \mpch$ with
periodic boundary  conditions. The particle mass is  $1.33$ x $10^{10}
\Msunh$.     Dark   matter   haloes    are   identified    using   the
friends-of-friends algorithm  \citep{Davis1985} with a  linking length
of 0.2 times the  mean inter-particle separation. Haloes obtained with
this linking length have a mean overdensity of 180 
\citep{Porciani2002}.  We consider only those  haloes which have  at 
least 20 particles or more.

To populate the dark matter haloes  with galaxies, we need to know the
number and the  luminosities of galaxies to be assigned  to each halo. 
Furthermore, we also need to assign phase space coordinates to each of
these galaxies. We use the  CLF described in \citet{Cacciato2008} for
the first purpose. The CLF is  a priori split into a contribution from
centrals and satellites, i.e.  $\Phi(L|M) = \philcm + \philsm$.  Here,
$\philcm \drm  L$ denotes the  conditional probability that a  halo of
mass  $M$ harbours  a central  galaxy  of luminosity  between $L$  and
$L+\drm  L$,  and $\philsm  \drm  L$  denotes  the average  number  of
satellites of  luminosity between $L$  and $L+\drm L$.  The parameters
that describe  the CLF are  constrained using the  luminosity function
\citep{Blanton2005} and  the luminosity dependence  of the correlation
length of galaxies \citep{Wang2007a} in SDSS.

Let us  consider a  halo of  mass $M$. The  luminosity of  the central
galaxy within  this halo is  sampled from the distribution  $\philcm$. 
The average number  of satellites that have a  luminosity greater than
$L_{\rm  min}=10^9 \Lsunh$  and reside  within haloes  of mass  $M$ is
given by
\begin{equation} \label{nsatm}
\avnsatm = \int_{L_{\rm min}}^{\infty} \philsm \drm L\,.
\end{equation}
We assume  Poisson statistics for the occupation  number of satellites
\citep{Kravtsov2004,Yang2005a,Yang2008}  and assign  $N_{\rm sat}$
galaxies to the halo where $N_{\rm sat}$ is drawn from
\begin{equation}
\label{poissondist}
 P(N_{\rm sat}|M) = \exp(-\mu) \frac{\mu^{N_{\rm sat}}}{N_{\rm sat}!}\,,
\end{equation}
with $\mu=\avnsatm$. The luminosities  of these satellite galaxies are
drawn from the distribution $\philsm$.

Phase space coordinates are assigned  to the galaxies in the following
manner. The central galaxy is assumed  to reside at rest at the centre
of the halo. Therefore, it has the same phase space coordinates as the
parent dark matter halo. We assume that the halo is spherical and that
the dark matter density distribution, $\rho(r|M)$, follows a universal
profile \citep{Navarro1997} given by
\begin{equation}\label{rhor}
 \rho(r|M) \propto \left( \frac{r}{r_s}\right)^{-1} \left( 1 +
\frac{r}{r_s} \right)^{-2}\,.
\end{equation}
Here, $r_s$ denotes a scale radius  which is specified in terms of the
virial  radius,  $r_{\rm  vir}$,  using the  concentration  parameter,
$c(M)=r_{\rm  vir}/r_s$. We use  the concentration-mass  relation from
\citet{Maccio2007}  appropriately modified for  our definition  of the
halo  mass. The  number  density distribution  of satellites,  $n_{\rm
  sat}(r|M)$, is given by
\begin{equation}\label{nsr}
n_{\rm sat}(r|M) \propto \left(
\frac{r}{ {\cal R} r_s}\right)^{-\gamma}
\left( 1 + \frac{r}{ {\cal R} r_s} \right)^{\gamma-3},
\end{equation}
where $\gamma$ represents the slope of the number density distribution
of  satellites,  as  $r  \rightarrow  0$  and ${\cal  R}$  is  a  free
parameter. For populating the  MGC, we adopt $\gamma={\cal R}=1$ which
implies that the satellites trace the dark matter density distribution
in  an  unbiased  manner.  The distribution,  $n_{\rm  sat}(r|M)$,  is
normalized such that
\begin{equation}
 \avnsatm = 4 \pi \, \int_0^{r_{\rm vir}} n_{\rm sat}(r|M) \, r^2 \,
\drm r \,.
\end{equation}
The radial coordinates  of the satellite galaxies with  respect to the
center  of  the  halo   are  sampled  from  the  distribution  $n_{\rm
  sat}(r|M)$. The satellite distribution around centrals is assumed to
be spherically  symmetric and random angular  coordinates are assigned
to  the  satellite  galaxies.  At  the  assigned  position  for  every
satellite  galaxy,  one  dimensional   velocities  are  drawn  from  a
Gaussian,
\begin{equation}
 f(v_j) = \frac{1}{\sqrt{2\,\pi}\sigma_{\rm sat}(r|M)} \exp \left[
-\frac{v_j^2}{2\sigma_{\rm sat}^2(r|M)}\right],
\label{drawsatvel}
\end{equation}
where  $v_j$  denotes the  relative  velocity  of  the satellite  with
respect to the central along axis $j$ and $\sigsat^2(r|M)$ denotes the
radial velocity  dispersion at a distance  $r$ from the  centre of the
halo. Here isotropy of orbits is assumed, i.e. the velocity dispersion
along the $j^{\rm th}$  axis, $\sigma^2_{j}$, equals $\sigsat^2(r|M)$. 
The  radial  velocity dispersion  within  the  halo  is determined  by
solving the Jeans equation
\begin{equation}\label{sigr}
\sigsat^2(r|M) = \frac{1}{n_{\rm sat}(r|M)}\int_r^{\infty}
n_{\rm sat}(r'|M)\frac{\partial \Psi}{ \partial r'}(r'|M) \rmd r'\,,
\end{equation}
where $\Psi(r'|M)$ is the gravitational potential and the radial
derivative represents the radial force given by
\begin{equation}\label{force}
 \frac{\partial \Psi}{ \partial r}(r|M) = \frac{4 \pi G}{r^2} 
\int_0^{r} \rho(r'|M) r'^2 \drm r'.
\end{equation}

We  assume  that  the  dark  matter  dominates  the  potential.  Using
Eqs.~(\ref{rhor})    and   (\ref{nsr})   in    Eqs.~(\ref{sigr})   and
(\ref{force}) gives
\begin{eqnarray}
\sigma^2_{\rm sat}(r|M) =  {c \, V^2_{\rm vir} \over {\cal R}^2
\mu(c)} \, \left({r \over {\cal R} r_s}\right)^\gamma \,
\left(1 + {r \over {\cal R} r_s}\right)^{3-\gamma} \, \nonumber \\
\int_{r/r_s}^{\infty} {\mu(x) {\rm d}x  \over
~(x / {\cal R})^{\gamma + 2}(1+ x / {\cal R} )^{3-\gamma}\,},
\label{sig1dm}
\end{eqnarray}
where $V_{\rm vir}$ is the circular velocity at $r_{\rm vir}$ and
\begin{equation}
\mu(x) = \int_0^x y (1+y)^{-2} \rmd y. %= \ln (1+x) - \frac{x}{1+x}
\end{equation}
The above expression for $\sigma^2_{\rm sat}(r|M)$ is used in the
distribution given by Eq.~(\ref{drawsatvel}) to assign velocities to
satellites. The entire procedure of assigning central and satellite
galaxies is repeated for all the dark matter haloes within the
simulation.

Our aim is  to construct a mock redshift survey that  mimics the SDSS. 
Therefore,  2x2x2 identical  galaxy-populated simulation  boxes (which
have periodic  boundary conditions) are stacked together.  A $(\rm RA,
DEC)$ coordinate frame  is defined with respect to  a virtual observer
at one  of the corners  of the stack.  The apparent magnitude  of each
galaxy is computed  according to its luminosity and  distance from the
observer. The line-of-sight (los) velocity of the galaxy is calculated
by adding  its peculiar velocity  to the velocity of  the cosmological
flow.  A random  velocity drawn  from a  Gaussian distribution  with a
dispersion of $35~\kms$ is further  added along the los to account for
the spectroscopic redshift errors present in the SDSS. The redshift as
seen  by  the  virtual  observer  is then  computed  using  the  total
velocity. We  only consider  galaxies with an  observed redshift  $z <
0.15$  and an  apparent  magnitude brighter  than  $17.77$. This  flux
limited  catalogue is  denoted  henceforth by  MOCKF  and has  289,500
galaxies above an absolute luminosity of $10^9 \Lsunhh$. MOCKF is used
in Appendix~\ref{sec:sample} to investigate potential selection biases
associated  with the  selection of  central galaxies.  In  addition to
MOCKF, we construct  a volume limited sample, MOCKV,  of galaxies that
lie in the redshift range 0.02~$\le z \le$~0.072 and have luminosities
greater than  $10^{9.5} \Lsunhh$. It  consists of 69,512  galaxies. In
what follows, we  use the volume limited sample  MOCKV to validate our
method for quantifying  the kinematics (Section~\ref{sec:veldisp}), to
validate  the  method to  infer  the  number  density distribution  of
satellites  (Section~\ref{sec:nsatprof}) and  finally to  confirm that
the mean  and scatter of the  MLR can be reliably  recovered using the
kinematics of satellites (Section~\ref{sec:model}). 

\section{Selection criteria}
\label{sec:selncrit}

Large  scale  galaxy redshift  surveys  such  as  the SDSS  allow  the
selection of  a statistically significant sample  of satellites. Since
the observed  galaxies cannot be  a priori classified as  centrals and
satellites,  it is  important to  use a  selection criterion  that can
correctly  identify central  galaxies and  the satellites  which orbit
around them. In this section, we describe the selection criterion that
we use to identify the central and satellite galaxies.

\begin{figure}
\centerline{\psfig{figure=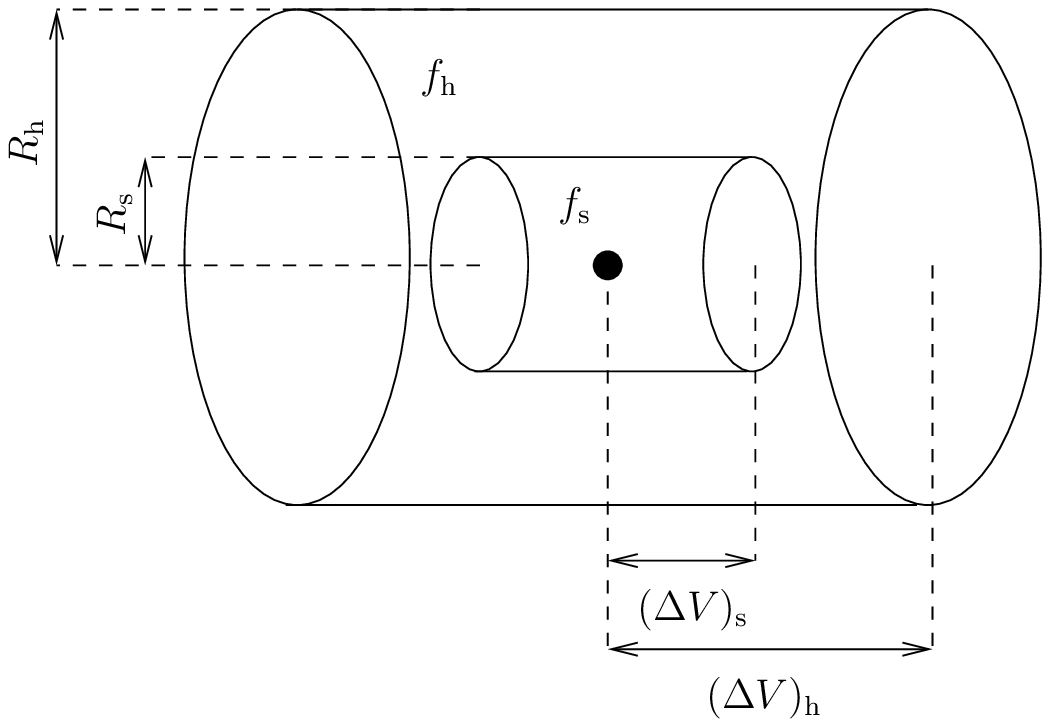,width=0.8 \hssize}}
\caption{ Schematic diagram of a selection criterion. Two coaxial 
  cylinders  are defined around  each galaxy  (represented by  a solid
  dot). The axis  is along the los while the face  of each cylinder is
  parallel to the plane of the sky.}
\label{fig:fig1}
\end{figure}

A galaxy is identified  to be a central if it is  at least $f_{\rm h}$
times  brighter than every  other galaxy  within a  cylindrical volume
specified  by  $R  < R_{\rm  h}$  and  $|\dv|  < (\dv)_{\rm  h}$  (see
Fig.~\ref{fig:fig1}). Here,  $R$ is the physical  separation from the
candidate central  galaxy projected  on the sky  and $\dv$ is  the los
velocity   difference.  Around  each   of  the   identified  centrals,
satellites  are those  galaxies that  are at  least $f_{\rm  s}$ times
fainter than their central galaxy  and lie within a cylindrical volume
specified  by $R  <  R_{\rm s}$  and  $|\dv| <  (\dv)_{\rm  s} $.  The
identification  of  the central  galaxies  depends  on the  parameters
$R_{\rm  h},(\dv)_{\rm h}$  and $f_{\rm  h}$, while  the  selection of
satellites  depends on  the parameters  $R_{\rm s},(\dv)_{\rm  s}$ and
$f_{\rm s}$. The  values of these parameters also  determine the level
of contamination of the sample  due to falsely identified centrals and
falsely  identified  satellites  (hereafter  interlopers).  The  false
identification of  centrals can be minimized by  choosing large values
of  $R_{\rm h},(\dv)_{\rm  h}$ and  $f_{\rm h}$  so that  the selected
central is the  dominant galaxy in a large volume.  On the other hand,
minimizing the  interlopers requires small  values of $R_{\rm  s}$ and
$(\dv)_{\rm s}$. A large value  of $f_{\rm s}$ further guarantees that
the selected satellites  are small and do not  dominate the kinematics
of  the  halo (i.e.  can  safely be  considered  as  test particles).  
Although stricter restrictions yield cleaner samples, they also reduce
the  sample size  significantly.  This makes  the velocity  dispersion
measurements   noisy.  Thus,   there   is  a   tradeoff  between   the
contamination level and the sample size.

\begin{table}
\caption{Selection Criteria}
\begin{center}
 \begin{tabular}{lcccccc}
   \hline
SC  & $R_{\rm h}$ & $(\dv)_{\rm h}$ & $f_{\rm h}$ & $R_{\rm s}$ &
$(\dv)_{\rm s}$ & $f_{\rm s}$ \\
   & Mpc/h & km/s & & Mpc/h & km/s & \\
\hline\hline
ITER  & $2.0$ & $4000$ & $1.0$ & $0.5$ & $4000$ & $1.0$ \\
    & $0.8 \sigma_{200}$ & $1000 \sigma_{200}$ & $1.0$ &
$0.15\sigma_{200}$ & $4000$ & $1.0$ \\
\hline
N08 & $1.0$ & $2400$ & $2.0$ & $0.4$ & $1200$ & $8.0$ \\
\hline
\end{tabular}
\end{center}

\medskip

\begin{minipage}{\hssize}
  The  parameters used  to specify  the inner  and the  outer cylinder
  around  a galaxy  for the  selection  criterion used  in this  paper
  (ITER) and the  one used in N08. The first row  for ITER denotes the
  parameters used in the first iteration, while the second row denotes
  the   parameters  used  in   subsequent  iterations.   The  velocity
  dispersion,  $\sigsat  $  in  units  of $200  \kms$  is  denoted  by
  $\sigma_{200}$  and  is  used   to  scale  the  cylinders  in  every
  iteration.
\end{minipage}
\label{tab:table1}
\end{table}

Most  authors have  chosen fixed  values for  the  selection criterion
parameters,  independent  of  the   luminosity  of  the  galaxy  under
consideration (McKay \etal 2002; Prada \etal 2003; Brainerd \& Specian
2003; N08). Since
brighter  centrals   on  average  reside  in   more  extended  haloes,
\citet{van den Bosch2004} advocated  an aperture which scales with the
virial radius  of the halo around  the galaxy. They  used an iterative
criterion  which  scales  the  cylindrical  aperture  based  upon  the
estimate  of the velocity  dispersion around  the central  after every
iteration.  In this  paper, we  also use  this iterative  criterion to
select  centrals  and  satellites.  The parameter  set  \{$R_{\rm  h},
(\dv)_{\rm h}$, $f_{\rm h}$, $R_{\rm s}, (\dv)_{\rm s}$, $f_{\rm s}$\}
that defines the inner and outer cylinders for the iterative criterion
(ITER) is  listed in Table~\ref{tab:table1}.  The first row  lists the
parameters  for the  first  iteration  while the  next  row lists  the
scaling of these parameters in the subsequent iterations. In short, we
proceed as follows:
\begin{enumerate}
\item Use  fixed values  of the aperture  size to select  centrals and
  satellites in the first iteration.
\item  Fit the  velocity dispersion  of the  selected satellites  as a
  function of  the central  galaxy luminosity, $\sigsat(\lc)$,  with a
  simple functional form (see Section~\ref{sec:veldisp_a}).
\item Select new centrals and  satellites by scaling the inner and the
  outer cylinders based on the estimate of the velocity dispersion.
\item Repeat (ii) and (iii)  until $\sigsat(\lc)$ has converged to the
  required accuracy.
\end{enumerate}
For step (iii), we adopt  the aperture scalings used in \citet{van den
  Bosch2004}. These aperture scalings  were optimised to yield a large
number of  centrals and  satellites, but at  the same time  reduce the
interloper  contamination.  The  values  chosen for  $R_{\rm  h}$  and
$R_{\rm  s}$ approximately  correspond to  $2$ and  $0.375$  times the
virial radius, $r_{\rm vir}$.

\section{Satellite kinematics}
\label{sec:veldisp}

In this  section, we  describe how to  measure and model  the velocity
dispersion-luminosity  relation,  $\sigma_{\rm  sat}(\lc)$, using  the
satellites  identified  by   the  selection  criterion.  The  relation
$\sigma_{\rm  sat}(\lc)$  can  be   measured  either  by  binning  the
satellites  by  central galaxy  luminosity  or  by  using an  unbinned
estimator. We use  the unbinned estimate after every  iteration of the
selection  criterion  to scale  the  selection  aperture. However,  to
quantify the kinematics of the  final sample of satellites, we use the
binned  estimator,  for  reasons  which  we  describe  further in the 
text.  In  the
following  subsections,  we  describe  the  unbinned  and  the  binned
estimators for $\sigma_{\rm sat}(\lc)$ and finally an analytical model
for the same.

\subsection{Unbinned Estimates}
\label{sec:veldisp_a}

We  use   a  maximum  likelihood  method  to   estimate  the  relation
$\sigma_{\rm sat}(\lc)$ from the  velocity information of the selected
satellites  after  every iteration  of  the  selection criterion.  Let
$\sigma_{200}$  denote $\sigsat  (\lc)$  in units  of  200 $\kms$  and
$L_{10}$  denote the  luminosity of  the  central galaxy  in units  of
$10^{10} \Lsunhh$. Following \citet{van den Bosch2004}, we parametrize
$\sigma_{200}$ as,
\begin{equation}
 \sigma_{200} (\log L_{10}) = a + b~(\log L_{10}) + c~(\log L_{10})^2.
\label{eq:sig200}
\end{equation}
Let $f_{\rm int}$ denote the  interloper fraction and assume that this
fraction is  independent of the  luminosity of the central  galaxy and
$\dv$. The probability for a selected satellite to have a los velocity
difference of $\dv \kms$ with respect to the central is then given by
\begin{equation}
 \pdv = \frac{f_{\rm int}}{2(\dv)_{\rm s}} + \frac{1-f_{\rm
int}}{\bar{\omega}} \exp\left[ - \frac{(\dv)^2}{2\sigma_{\rm eff}^2}
\right],
\end{equation}
where,  $\sigma_{\rm  eff}   =  [\sigma_{\rm  sat}^2  +  \sigma^2_{\rm
  err}]^{1/2} $  is the effective velocity dispersion  in the presence
of the redshift errors and the factor
\begin{equation}
 \bar{\omega} = \sqrt{2 \pi}\sigma_{\rm eff}\,{\rm erf}\left[
\frac{(\dv)_{\rm s}}{\sqrt{2}\sigma_{\rm eff}} \right],
\end{equation}
is  such that  the  $\pdv$ is  properly  normalized to  unity. In  our
attempt to mimic SDSS, we have  added a Gaussian error of $35 \kms$ to
the velocity of each galaxy  in the mock catalog. Therefore, the error
on  the  velocity difference,  $\dv$,  of  the  central and  satellite
galaxies is $\sigma_{\rm err}  = \sqrt{2}\,\times\,35 \kms$ which adds
in quadrature to $\sigma_{\rm sat}$ to yield $\sigma_{\rm eff}$.

We  use Powell's  direction  set method  to  determine the  parameters
($a,~b,~c,~f_{\rm  int}$) that  maximize  the likelihood  ${\cal L}  =
\sum_i  \ln[\pdv]_i$, where  the summation  is over  all  the selected
satellites.  This  yields  a  continuous  estimate  of  $\sigsat(\lc)$
without the  need to  bin the los  velocity information  of satellites
according to the  luminosity of the central galaxy.  The parameter set
($a,~b,~c$) fitted in  the last but one iteration  determines the size
of the  apertures used to select  the final sample  of satellites. The
values of these parameters for  the samples investigated in this paper
are listed in Table~\ref{tab:table2}.

\begin{table}
\caption{Parameters for the Selection Criterion}
\begin{center}
 \begin{tabular}{lccc}
   \hline
Sample & a & b & c
\\
\hline\hline
MOCKV &  2.06 &  0.45 & 0.25 \\
MOCKF &  2.05 &  0.50 & 0.23 \\
SDSSV &  2.20 &  0.38 & 0.33 \\
\hline
\end{tabular}
\end{center}
\medskip

\begin{minipage}{\hssize}
  The   parameters    used   in   Eq.~(\ref{eq:sig200})    to   define
  $\sigma_{200}$ as  a function of the  luminosity of a  galaxy in the
  final iteration for samples  MOCKV, MOCKF and for the volume-limited
  sample from the SDSS, denoted by SDSSV.
\end{minipage}
\label{tab:table2}
\end{table}
\begin{figure}
\centerline{\psfig{figure=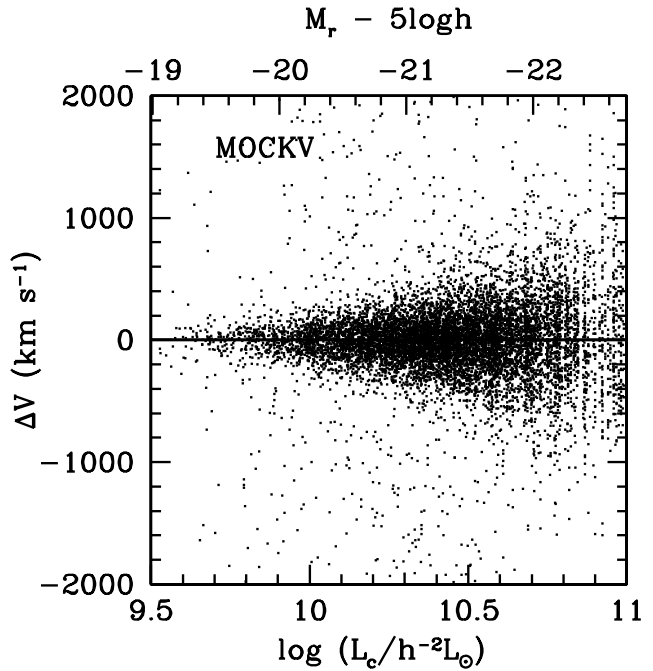,width=\hssize}}
\caption{Scatter plot of the velocity difference, $\dv$, between
  the  satellites and  their centrals  as  a function  of the  central
  galaxy  luminosity. The  satellites  were obtained  by applying  the
  iterative selection criterion to MOCKV.}
\label{fig:fig2}
\end{figure}

\subsection{Binned Estimates}
\label{sec:veldisp_b}

We  use a binned  estimator to  quantify the  kinematics of  the final
sample  of satellites.  The binned  estimator allows  us to  relax the
simplistic  assumption of $f_{\rm  int}$ being  independent of  $\lc$. 
More importantly, the binned estimator allows us, in a straightforward
manner,  to  measure   $\sigma_{\rm  sat}(\lc)$  using  two  different
weighting  schemes  -  satellite-weighting  and  host-weighting.  Most
studies in the literature have used one of these two weighting schemes
to infer the mean of the MLR. However, as demonstrated in Paper I, the
mean of  the MLR inferred from  the velocity dispersion in  any one of
these  two schemes is  degenerate with  the scatter  in the  MLR. This
degeneracy can be broken by modelling the velocity dispersions in both
schemes  {\it simultaneously}.  In  what follows,  we briefly  explain
these two weighting schemes in  turn and then verify that the velocity
dispersions in both schemes can be accurately recovered from the MGC.

\begin{figure*}
\centerline{\psfig{figure=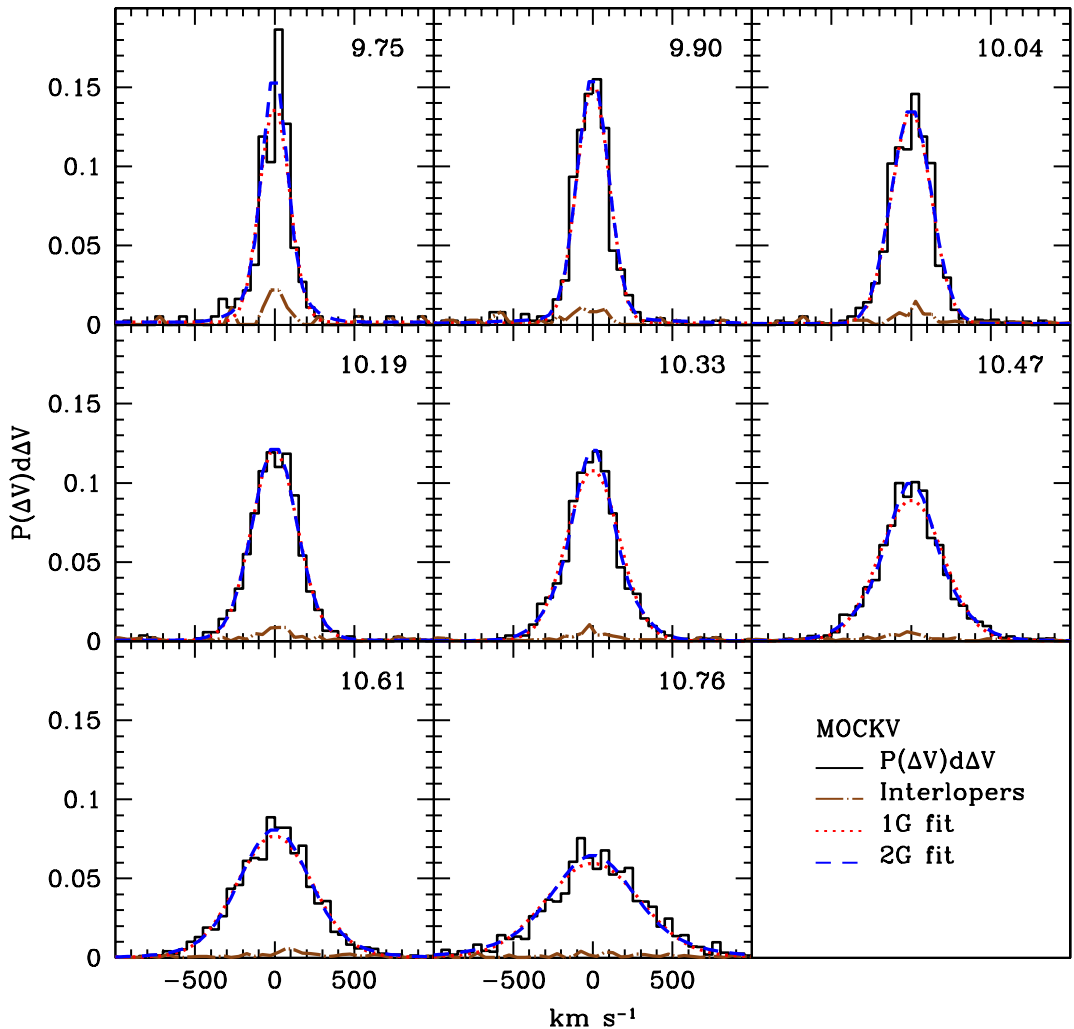,width=0.8 \hdsize}}
\caption{ The satellite-weighted $\pdv$ distributions of satellites
  around centrals selected in  several luminosity bins from MOCKV. The
  average  $\log(\lc/\Lsunh)$ for each  bin is  indicate at  the upper
  right  corner of  every panel.  The (brown)  dot-dashed line  at the
  bottom of  each distribution shows  the contamination of  the $\pdv$
  distributions  due  to  the  interlopers.  The  (red)  dotted  lines
  indicate  the single  Gaussian  fits, the  (blue)  dashed lines  the
  double Gaussian fits.}
\label{fig:fig3}
\end{figure*}

To   measure   the   velocity   dispersion  of   satellites   in   the
satellite-weighting scheme,  we obtain the  distribution of velocities
of the satellites, $\pdv$, with  respect to their centrals for several
bins of central  galaxy luminosity (see Fig.~\ref{fig:fig3}). Each
bin has a width $\Delta \log [\lc]=0.15$. In this scheme, the centrals
that have a larger number of satellites clearly contribute more to the
$\pdv$  distribution  than  those  which  have  a  smaller  number  of
satellites.  Therefore,   the  resulting   scatter  in  $\pdv$   is  a
satellite-weighted  average  of the  velocity  dispersions around  the
stacked  centrals  (see  Paper  I  for  a  detailed  discussion).  The
dispersion  obtained  using  this  scheme  is  denoted  henceforth  by
$\sigsw$.

One has  to undo the  satellite-weighting described above in  order to
measure   the   host-weighted  velocity   dispersion.   This  can   be
accomplished   by   introducing   a   weight   $w=N^{-1}$   for   each
central-satellite  pair  while  constructing the  $\pdv$  distribution
\citep{van den  Bosch2004, Conroy2007, Becker2007}.  Here, $N$ denotes
the   number  of   satellites  selected   around  the   central  under
consideration. Therefore, in this scheme each central receives a total
weight of unity irrespective of the number of satellites it hosts. The
scatter  in this  weighted  $\pdv$ distribution  is the  host-weighted
velocity dispersion and is denoted henceforth by $\sighw$.

The procedure to obtain the scatter in the $\pdv$ distributions is the
same for both the satellite-weighted and the host-weighted case.  This
procedure  must account for  the interlopers  and the  redshift errors
present in MOCKV.  In what follows, we illustrate  this procedure only
for the satellite-weighted case.

Fig.~\ref{fig:fig2}   shows  the   scatter  plot   of  velocity
difference  $\dv$ of  the selected  satellites and  the centrals  as a
function  of the  luminosity of  the centrals.  The satellite-weighted
$\pdv$ distributions of the satellites selected from MOCKV for several
central  luminosity   bins  are  shown   in  Fig.~\ref{fig:fig3}.  
Dot-dashed lines  show the  contamination of the  $\pdv$ distributions
due  to interlopers  and  are barely  visible  at the  bottom of  each
distribution. This  confirms the  claims in \citet{van  den Bosch2004}
that the  iterative criterion yields  a small fraction  of interlopers
with  a weak  dependence  on $\lc$  and  that the  interlopers can  be
modelled  as  a constant  contribution  to  the velocity  distribution
independent of $\dv$.

\begin{figure*}
\centerline{\psfig{figure=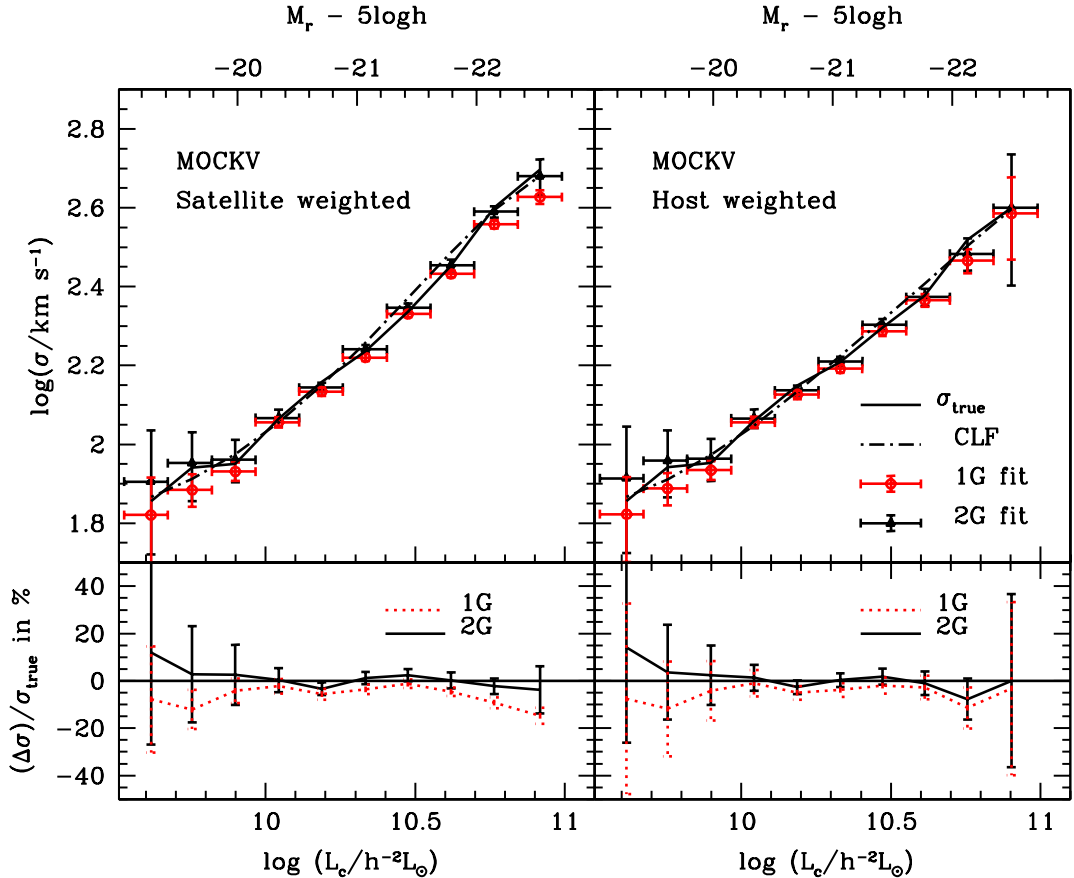,width=0.8\hdsize }}
\caption{Upper panels show the satellite-weighted and the
  host-weighted velocity  dispersions recovered from  MOCKV. The (red)
  circles show values  recovered from a single Gaussian  fit while the
  (black) triangles show those from the double Gaussian fit. The solid
  line shows  the variance of  the true satellites and  the dot-dashed
  line shows  the analytical prediction  using the halo  occupation
  statistics of
  centrals  from  the  MGC.  The  bottom panel  plots  the  percentage
  deviation of the  single and double Gaussian fits  from the variance
  of the true satellites.}
\label{fig:fig4}
\end{figure*}

A simple  way to estimate the  scatter of a $\pdv$  distribution is to
fit a Gaussian plus a constant model given by
\begin{equation}
 \pdv = a_0 + a_1 {\rm exp}\left[\frac{-(\dv)^2}{2
\sigma_{\rm eff}^2}\right].
\end{equation}
Here, $a_0$  denotes the constant  (with respect to  $\dv$) interloper
background,   $a_1$  is   the  normalization   of  the   Gaussian  and
$\sigma_{\rm eff}$ is the effective  dispersion in the presence of the
redshift errors.

The velocity dispersion obtained using a single Gaussian plus constant
model fit can be systematically affected if the $\pdv$ distribution is
intrinsically non-Gaussian. \citet{Diaferio1996} demonstrated that the
velocity distribution  can be non-Gaussian  partly due to  mass mixing
(which is the result of  stacking haloes of different mass) and partly
due to  the unrelaxed  state of a  halo. The  second moment of  such a
non-Gaussian distribution can be estimated with a double Gaussian plus
a constant model \citep{Becker2007} given by
\begin{equation}
\pdv = a_0 + a_1 {\rm exp}\left[\frac{-(\dv)^2}{2
\sigma_{1}^2}\right] + a_2 {\rm exp}\left[\frac{-(\dv)^2}{2
\sigma_{2}^2}\right].
\end{equation}
The scatter, $\sigma_{\rm eff}$, in this case is such that
\begin{equation}
 \sigma_{\rm eff}^2 =
\frac{a_1\sigma_1^3+a_2\sigma_2^3}{a_1\sigma_1+a_2\sigma_2}
=\sigsw^2 + \sigma_{\rm err}^2.
\end{equation}
The dotted  and the dashed  lines in Fig.~\ref{fig:fig3}  show the
single  and   double  Gaussian   fits  to  the   $\pdv$  distributions
respectively.

Fig.~\ref{fig:fig4}  shows the  velocity  dispersions obtained
from  the satellite-weighted  $\pdv$ distributions  in the  upper left
panel and  those obtained from the  host-weighted $\pdv$ distributions
in the upper right panel.  The (red) circles and the (black) triangles
indicate the  single and the double Gaussian  fits respectively. Since
the true  satellites of centrals  selected from MOCKV are  known, they
can be used to judge  the goodness of the fits. The satellite-weighted
and  the host-weighted  velocity  dispersions of  the true  satellites
(among  the satellites  selected  using the  iterative criterion)  are
obtained using
\begin{equation}
 \sigma^2_{\rm true} = { \sum_{j=1}^{N_{\rm c}} \sum_{i=1}^{N_j}
w_{ij} (\dv)_{ij}^2 \over \sum_{j=1}^{N_{\rm c}} \sum_{i=1}^{N_j}
w_{ij} } - \sigma_{\rm err}^2.
\end{equation}
Here, $N_{\rm c}$  denotes the number of true  centrals, $N_j$ denotes
the  number  of  true  satellites  of the  $j^{\rm  th}$  central  and
$(\dv)_{ij}$ denotes  the los velocity difference of  the $j^{\rm th}$
central  with  respect  to  its  $i^{\rm th}$  satellite.  The  weight
$w_{ij}=1$ for  the satellite-weighted case  and $w_{ij}=N_j^{-1}$ for
the host-weighted  case. The  true velocity dispersions  thus obtained
are shown as solid curves in Fig.~\ref{fig:fig4}.

The bottom  panels of Fig.~\ref{fig:fig4}  show the percentage
deviation of  both the  single and the  double Gaussian fits  from the
velocity dispersions  of the true satellites. The  single Gaussian fit
(the  dotted line)  underestimates the  dispersions  systematically by
about $5-10$\%. The double Gaussian  fit (the solid line) on the other
hand  gives  an  unbiased  estimate  of  both  velocity  dispersions.  
Therefore,  in  what follows,  we  use  the  double Gaussian  fit  for
measuring both  the satellite-weighted and  the host-weighted velocity
dispersions (cf. Becker \etal 2007).

\subsection{Analytical Estimates}
\label{sec:veldisp_c}

We   now  compare   the   velocity  dispersions   obtained  from   the
satellite-weighted and  the host-weighted schemes  to their analytical
expectation values. As detailed in Paper I, the satellite-weighted and
the host-weighted  velocity dispersions depend on  the distribution of
halo masses  of central galaxies specified by  $\pmlc$. The analytical
expressions describing the velocity  dispersion in these two weighting
schemes are
\begin{equation}
\sigsw^2(L_c) =
\frac{ \int_{0}^{\infty} \,
\pmlc \, \avnsat_{{\rm ap},M} \, \avgsigsatsq_{{\rm ap},M} \,
\drm M }{ \int_{0}^\infty \,
\pmlc \, \avnsat_{{\rm ap},M} \drm M } \,,
\label{sweqnvol} 
\end{equation}
\begin{equation}
\sighw^2(L_c) =
\frac{ \int_{0}^{\infty} \, \pmlc \,
{\cal P}(\avnsat_{{\rm ap},M}) \,
\avgsigsatsq_{{\rm ap},M} \, \drm M} { \int_{0}^\infty \, \pmlc \,
{\cal P}(\avnsat_{{\rm ap},M}) \, \drm M }\,.
\label{hweqnvol}
\end{equation}
Here,  the  average number  of  satellites  and  the average  velocity
dispersion of satellites, within the aperture $R_{\rm s}$ in a halo of
mass    $M$,    are   denoted    by    $\avnsat_{{\rm   ap},M}$    and
$\avgsigsatsq_{{\rm  ap},M}$ respectively.  The  number of  satellites
within the aperture, $\avnsat_{{\rm ap},M}$, is given by
\begin{equation}\label{nsatap}
 \avnsat_{{\rm ap},M} = 4 \pi \int_0^{R_{\rm s}} R \, \rmd R
\int_{R}^{r_{\rm vir}} n_{\rm sat}(r|M) \, {r \rmd r \over \sqrt{r^2
- R^2}}\,.
\end{equation}
Assuming the  velocity dispersion of  satellites to be  isotropic, the
velocity dispersion, $\avgsigsatsq_{{\rm  ap},M}$, can be expressed as
the   average  of   the  radial   velocity   dispersion,  $\sigma_{\rm
  sat}^2(r|M)$ (see Eq.~\ref{sig1dm}), over the aperture $R_{\rm s}$,
\begin{eqnarray}\label{sigaper}
\avg{\sigma_{\rm sat}^2}_{{\rm ap},M} &=& { 4 \pi \over \avnsat_{{\rm
ap},M}} \int_0^{R_{\rm s}} R \, {\rm d}R \nonumber \\
& & \int_{R}^{r_{\rm vir}} n_{\rm sat}(r|M) \, \sigma_{\rm
sat}^2(r|M) \, {r \rmd r\over \sqrt{r^2 - R^2}}\,.
\end{eqnarray}
Note that, when measuring  the host-weighted velocity dispersions only
satellites of those  centrals that have at least  one satellite within
the search aperture are used. The fraction of such centrals is denoted
by ${\cal  P}(\avnsat_{{\rm ap},M})$ and  is given by  the probability
that a halo of mass $M$, which on average hosts $\avnsat_{{\rm ap},M}$
satellites within  the aperture $R_{\rm  s}$, has $N_{\rm sat}  \ge 1$
within the aperture. Therefore,
\begin{eqnarray}
{\cal P}(\avnsat_{{\rm ap},M}) &\equiv&  P(N_{\rm sat} \ge 1)
\nonumber \\
                               &=& 1 - P(N_{\rm sat} = 0)  
\nonumber \\
                               &=& 1 - \exp[-\avnsat_{{\rm ap},M}]
\label{poisson}
\end{eqnarray}
In the  last equality  we have assumed  that the  satellite occupation
numbers (cf.  Eq.~\ref{poissondist}) follow Poisson  statistics, which
is supported by numerical simulations \citep{Kravtsov2004} and by
results from group catalogs based on SDSS \citep{Yang2005a,Yang2008}.
The factor ${\cal  P}(\avnsat_{{\rm ap},M})$ is not considered in the 
analytical estimate in the satellite-weighting scheme as haloes with 
zero satellites, by definition, contribute zero weight.

\begin{figure*}
\centerline{\psfig{figure=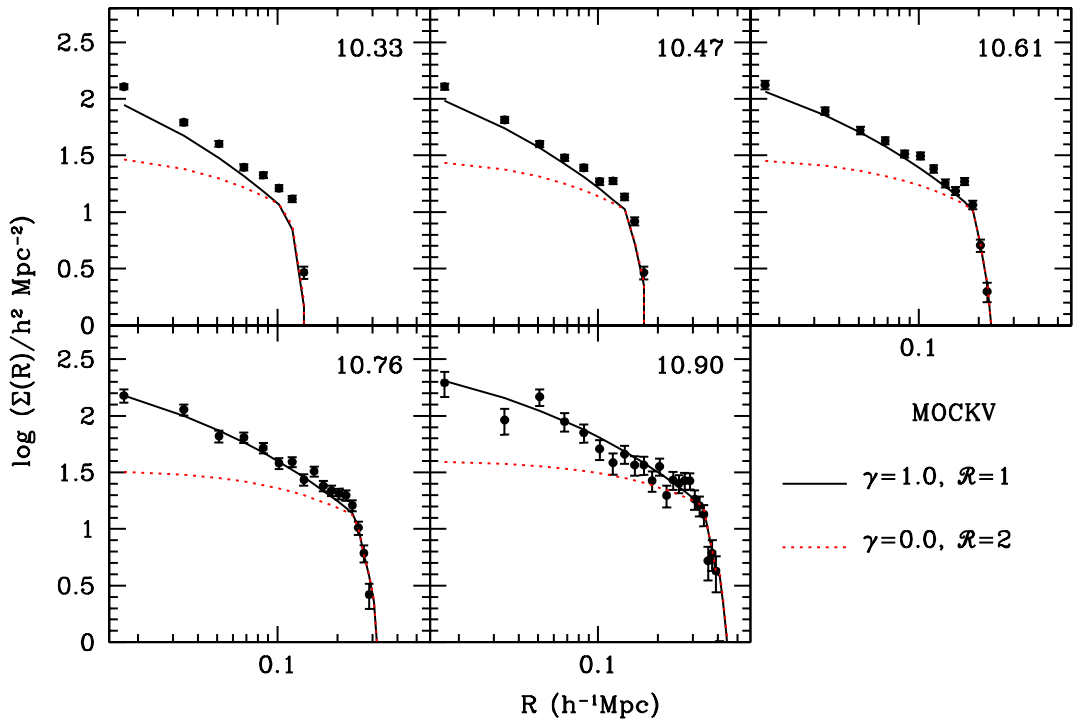,width=0.8 \hdsize}}
\caption{The projected number density distributions of the satellites
  selected from  MOCKV as  a function of  the projected radius  in the
  brightest  central  luminosity bins.  The  errorbars assume  Poisson
  statistics for  the number  of satellites in  each bin.  The (black)
  solid lines indicate the  analytical predictions and assume that the
  satellite  number  density  distribution  follows  the  dark  matter
  distribution in an unbiased manner, i.e. ${\cal R}=1$ and $\gamma=1$
  in eq.~(\ref{nsr}). For comparison,  the (red) dotted lines show the
  analytical predictions that assume ${\cal R}=2$ and $\gamma=0$.}
\label{fig:fig5}
\end{figure*}

From the above analytical description, it is clear that the analytical
estimates for the velocity dispersions require the knowledge of
\begin{itemize}
\item the density distribution of dark matter haloes, $\rho(r|M)$
 \item  the   number  density  distribution   of  satellites,  $n_{\rm
     sat}(r|M)$
\item  the  halo  occupation statistics of  centrals,  $\pmlc$,  and
     the halo occupation number of satellites, $\avnsatm$.
\end{itemize}

We assume that the density distribution of dark matter haloes is given
by  Eq.~(\ref{rhor})  and  that  the number  density  distribution  of
satellites is given by  Eq.~(\ref{nsr}) with $\gamma={\cal R}=1$.  The
halo occupation statistics of centrals, $\pmlc$ is given by
\begin{equation}
 \pmlc = { \Phi_{\rm c}(\lc|M) n(M) \over \int \Phi_{\rm c}(\lc|M)
n(M) \drm M }
\end{equation}
where $\Phi_{\rm c}(\lc|M)$ is  the conditional luminosity function of
central galaxies and  $n(M)$ is the halo mass  function. The number of
satellite   galaxies   in  a   halo   of   mass   $M$  is   given   by
Eq.~(\ref{nsatm}).  We  adopt the $\Phi_{\rm  c}(\lc|M)$ and $\philsm$
that were used in  Section~\ref{sec:mockcat} to populate the MGC. With
this  input, we  compute  the analytical  estimates  for the  velocity
dispersions  of   satellites  as   a  function  of   luminosity  using
Eqs.~(\ref{sweqnvol}) and (\ref{hweqnvol}).  The results thus obtained
are  shown  as  dot-dashed  curves  in  the  corresponding  panels  of
Fig.~\ref{fig:fig4}. Overall  the agreement with  the velocity
dispersions  obtained from  the satellites  in the  MGC is  very good,
except at intermediate luminosities where the analytical estimates are
$\sim 5$ percent higher  than $\sigma_{\rm true}$. This indicates that
the central galaxies selected from the  MGC  do  not properly  sample 
the  full
$\pmlc$. This can have two  reasons: (i) a systematic problem with the
criterion used to select central galaxies, or (ii) cosmic variance due
to the finite volume probed by MOCKV.  As we demonstrate in Appendix~A
our iterative  criterion accurately  samples the true  $\pmlc$, except
for the  fact that it misses  the haloes of those  centrals which have
zero satellites.   However, this sampling  effect is accounted  for in
our analytical model via  Eq.~(\ref{poisson}).  In fact detailed tests
show  that  the  discrepancies  between $\sigma_{\rm  true}$  and  our
analytical estimates are entirely due to cosmic variance in the MGC.

In Appendix~{\ref{sec:sample}}, we also show that the strict selection
criteria, that have been abundantly  used in the literature, lead to a
sample of central galaxies that  is biased to reside in relatively low
mass haloes.   Consequently, the resulting MLR of  central galaxies is
similarly biased, and has to be interpreted with great care.

\section{Number Density Distribution of Satellites}
\label{sec:nsatprof}

As  described  above, the  number  density  distribution of  satellite
galaxies,  $n_{\rm sat}(r|M)$,  is a  necessary input  to analytically
compute  the  velocity   dispersions.  The  projected  number  density
distribution  of  satellites, $\Sigma(R|\lc)$,  around  centrals of  a
given  luminosity, directly  reflects the  functional form  of $n_{\rm
  sat}(r|M)$.  The   distribution  $\Sigma(R|\lc)$  can   be  directly
measured  by  combining the  satellites  around  centrals  of a  given
luminosity, $\lc$,  chosen by the selection criterion.  However, it is
necessary to  first assess the impact of  the interloper contamination
on the measurement of $\Sigma(R|\lc)$,  for which we again make use of
the satellite sample selected from MOCKV.

Fig.~\ref{fig:fig5}  shows,  for  the five  brightest  luminosity
bins, the azimuthally  averaged projected number density distributions
of  the satellites  selected  from MOCKV.  The  errorbars reflect  the
Poisson  noise on the  number of  satellites in  each radial  bin. The
abrupt cutoff at large $R$ is an artefact due to the parameter $R_{\rm
  s}$ in the selection criterion which describes the maximum projected
radius within which satellites get selected.  Note that, since $R_{\rm
  s}$  depends   upon  the   luminosity  of  central   galaxies  under
consideration,  this  cutoff  shifts  to larger  $R$  with  increasing
central galaxy luminosity.

The  projected  number   density  distribution  of  satellites  around
centrals  stacked according  to luminosity,  $\Sigma (R|\lc)$,  can be
analytically expressed as,
\begin{equation}
\label{satproflc}
\Sigma(R|\lc) = \frac{\int \, \pmlc \, \Sigma (R|M) \, \drm M
}{\int \, \pmlc \, {\cal P}(\avnsatm_{{\rm ap},M}) \, \drm M }.
\end{equation}
Here, $\Sigma (R|M)$ is the projection of $n_{\rm sat}(r|M)$ along the
line-of-sight and is given by
\begin{equation}
\Sigma (R|M) =  \int_{R}^{r_{\rm vir}} \frac{n_{\rm sat}(r|M)\,  2
r \, \drm r}{\sqrt{r^2-R^2}},
\end{equation}
Using  $n_{\rm   sat}(r|M)$  given  by   Eq.~(\ref{nsr})  with  ${\cal
  R}=\gamma=1$  and   the  true  $\pmlc$   present  in  the   MGC,  we
analytically  compute  the  expected  number density  distribution  of
satellites around centrals  of a given luminosity. The  solid lines in
Fig.~\ref{fig:fig5}   show  the   results   of  this   analytical
expectation. The small differences between the measured and
the analytically obtained
distributions are due to the interlopers in the sample. However,
the differences
become negligible in the brighter luminosity bins. For comparison, the
(red) dotted lines show  the expected $\Sigma(R|\lc)$ for ${\cal R}=2$
and  $\gamma=0$.  This  shows  that  the  parameters  ${\cal  R}$  and
$\gamma$,  that  characterize   the  number  density  distribution  of
satellites,  can  be  inferred   from  the  projected  number  density
distributions of the selected satellites.

\section{Mass-Luminosity Relationship}
\label{sec:model}

In the previous  sections, using a mock catalog,  we have demonstrated
that  the satellite-weighted  velocity dispersions,  the host-weighted
velocity dispersions and the projected number density distributions of
satellites  around centrals  of  a given  luminosity  can be  reliably
measured starting from a volume  limited redshift catalog of galaxies. 
Next,  we  attempt to  infer  the MLR  of  central  galaxies from  the
velocity  dispersions  measured  from  MOCKV.  The aim  is  to  invert
Eqs.~(\ref{sweqnvol})   and   (\ref{hweqnvol})   which  describe   the
dependence of the velocity dispersions on the MLR of central galaxies.
In addition to  the velocity dispersions, we also  measure the average
number   of   satellites   per   central  of   a   given   luminosity,
$\avnsat(\lc)$,  and  use this  as  a  constraint.  The dependence  of
$\avnsat(\lc)$ on the MLR of central galaxies is given by
\begin{equation}
\avnsat(L_c) =
\frac{ \int_{0}^{\infty} \pmlc
\avnsat_{{\rm ap},M} \drm M} { \int_{0}^\infty \pmlc  {\cal
P}(\avnsat_{{\rm ap},M}) \drm M }.
\label{nsateqnvol}
\end{equation}
In this section, we first describe  the model we use to infer the mean
and the scatter of the  MLR from the observables $\sigsw,\,\sighw$ and
$\avnsat$. Next, we  use this model to infer the  mean and the scatter
of the MLR in the MGC and  compare it to the true relations present in
the MGC.

\begin{figure*}
\centerline{\psfig{figure=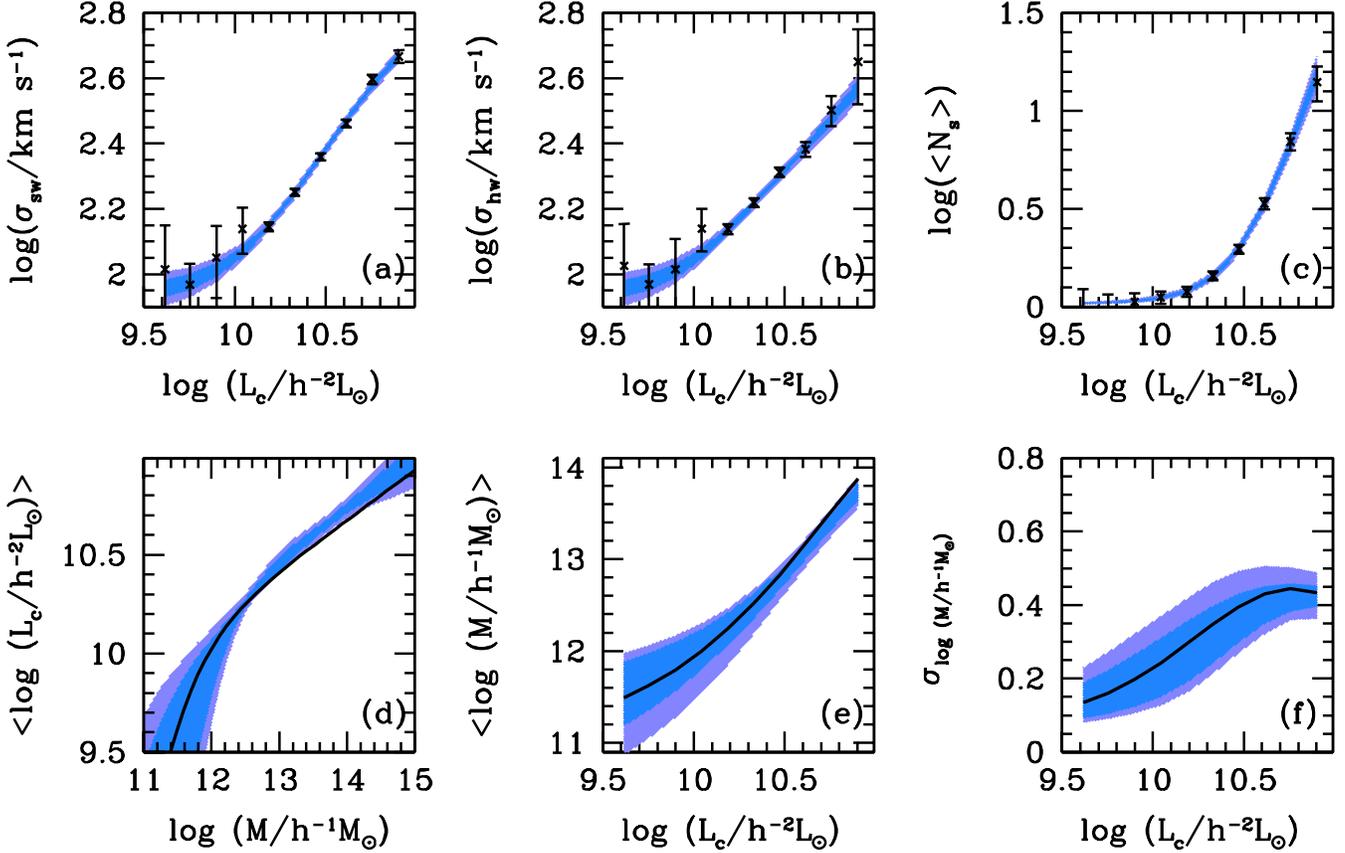,width= \hdsize}}
\caption{Crosses with errorbars in the upper panels denote the data
  used   to   constrain   the    Monte   Carlo   Markov   Chain;   the
  satellite-weighted   velocity   dispersions   in  panel   (a),   the
  host-weighted  velocity dispersions  in  panel (b)  and the  average
  number  of  satellites  per  central  in panel  (c).  The  relations
  recovered from  the MCMC  analysis are shown  in the  bottom panels;
  $\avgloglc$ in panel (d), $\avglogm$  in panel (e) and $\siglogm$ in
  panel (f).  In each  panel, the blue  and purple colours  denote the
  68\% and  the 95\% confidence levels.  The solid lines  in the lower
  panels denote the true relations present in MOCKV.}
\label{fig:fig6}
\end{figure*}

\subsection{The Model}
\label{sec:model_a}

As mentioned earlier,  the analytical computation of $\sigsw,\,\sighw$
and $\avnsat$  requires the knowledge  of the density  distribution of
dark matter haloes, the  number density distribution of satellites and
the halo  occupation statistics of  centrals and satellites.  We
assume that the
density  distribution of dark  matter haloes  follows the  NFW profile
given  by Eq.~(\ref{rhor}).  For  the number  density distribution  of
satellites  within a  halo of  mass  $M$, $n_{\rm  sat}(r|M)$, we  use
Eq.~(\ref{nsr})    with    ${\cal    R}=\gamma=1$.   As    shown    in
Section~\ref{sec:nsatprof}, the projected number density distributions
of satellites  selected from MOCKV is consistent  with this analytical
expression. Next, we describe our model for the halo occupation
statistics of the
centrals,  specified  by $\pmlc$,  and  the  satellites, specified  by
$\avnsatm$.
\begin{table}
\caption{MOCKV: Parameters recovered from the MCMC}
\begin{center}
 \begin{tabular}{ccccc}
   \hline
Parameter & Input & 16\% & 50\% & 84\%
\\
\hline\hline
$\log (L_0)$ & 9.93 &  9.64 &   10.01 & 10.32\\
$\log (M_1)$ & 11.04& 10.48 &   11.28 & 11.69\\
$\gamma_2$   & 0.25 & 0.18  &    0.26 & 0.32\\
$\sigcen$    & 0.14 & 0.13  &    0.15 & 0.17\\
\hline
\end{tabular}
\end{center}
\medskip
\begin{minipage}{\hssize}
  The  input parameters  that describe  $\plcm$ are
  compared to the  $16^{\rm th}$, $50^{\rm th}$ and  the $84^{\rm th}$
  percentiles  of   the  corresponding  distributions   of  parameters
  obtained from the MCMC.
\end{minipage}
\label{tab:table3}
\end{table}

The distribution $\pmlc$ is related to the complementary distribution,
$\plcm$, by Bayes' theorem
\begin{equation}
 P(M|\lc) = \frac{ n(M) \plcm } {\int n(M) \plcm \drm M },
\end{equation}
where $n(M)$ is the halo mass function. We follow \citet{Cacciato2008}
and parametrize the distribution $\plcm$ as a lognormal in $\lc$,
\begin{equation}
 \plcm \drm \lc = {\log(e) \over \sqrt{2 \pi} \sigcen}
\exp\left[ - {(\log [\lc/\tilde{\lc}])^2 \over 2 \sigcen^2 } \right]
\frac{\drm \lc}{\lc}.
\end{equation}
Here  $\log  \tilde{\lc}  (M)$  denotes  the  mean  of  the  lognormal
distribution and $\sigcen$ is the scatter in this distribution. We use
four parameters  to specify the relation $\tilde{\lc}(M)$:  a low mass
end  slope,   $\gamma_1$,  a  high  mass  end   slope,  $\gamma_2$,  a
characteristic mass scale, $M_1$, and a normalisation, $L_0$;
\begin{equation}
\tilde{\lc} = L_0  \frac{ \left(M/M_1\right)^{\gamma_1} }{ [1
+ \left( M/M_1\right)]^{\gamma_1-\gamma_2}}.
\end{equation}
We assume the  scatter $\sigcen$ to be independent of  mass. We do not
explore  the faint  end slope,  $\gamma_1$,  in our  analysis as  the
velocity dispersions  at the faint end  are very uncertain  due to low
number statistics. Instead, we keep  it fixed at $3.273$, which is the
value obtained  from the analysis  of the abundance and  clustering of
galaxies \citep[see][]{Cacciato2008}. This parametrization is
motivated  by results of \citet{Yang2008}  who measure the conditional
luminosity function from the SDSS group catalogue described in
\citet{Yang2007}.

We model the  satellite occupation number, $\avnsatm$, as  a power law
distribution, given by
\begin{equation}
 \avnsatm = N_s \left(\frac{M}{10^{12} \Msunh}\right)^{\alpha}\,,
\end{equation}
which  adds two more  parameters ($N_s,\alpha$).  Thus, in  total, our
model has  six free parameters  ($\sigcen, L_{0}, M_1,  \gamma_2, N_s,
\alpha$).  Given  these  parameters  and  the  radial  number  density
distribution of satellites (specified by ${\cal R}$ and $\gamma$), the
velocity dispersions  $\sigsw(\lc)$ and  $\sighw(\lc)$ as well  as the
number of satellites  per central, $\avnsat(\lc)$ in an  aperture of a
given    size   can    be   computed    using   Eqs.~(\ref{sweqnvol}),
(\ref{hweqnvol}) and  (\ref{nsateqnvol}) and compared  to the measured
values.  Crosses  with  errorbars  in  Panels  (a),  (b)  and  (c)  of
Fig.~\ref{fig:fig6} show $\sigsw$, $\sighw$ and $\avnsat$ as a
function of the luminosity of the central obtained from MOCKV.  We use
these measurements to constrain the six free parameters of our model.

\subsection{Monte-Carlo Markov Chain }

To  determine the posterior  probability distributions  of the  6 free
parameters  in  our  model,   we  use  the  Monte-Carlo  Markov  Chain
(hereafter MCMC) technique. The MCMC is a chain of models, each with 6
parameters. At any point in the chain, a trial model is generated with
the 6 free parameters  drawn from 6 independent Gaussian distributions
which  are  centred  on   the  current  values  of  the  corresponding
parameters.  The chi-squared statistic,  $\chi^2_{\rm try}$,  for this
trial model, is calculated using
\begin{equation}
 \chi^2_{\rm try} = \chi^2_{\rm sw} + \chi^2_{\rm hw} + \chi^2_{\rm
ns}\, ,
\end{equation}
with
\begin{eqnarray}
 \chi^2_{\rm sw} &=& \sum_{i=1}^{10} \left[{\sigsw(\lc[i]) -
\hat{\sigma}_{\rm sw}(\lc[i]) \over \Delta \hat{\sigma}_{\rm
sw}(\lc[i])} \right]^2 \,, \\
\chi^2_{\rm hw} &=& \sum_{i=1}^{10} \left[{\sighw(\lc[i])
- \hat{\sigma}_{\rm hw}(\lc[i]) \over \Delta \hat{\sigma}_{\rm
hw}(\lc[i])} \right]^2 \, , \\
\chi^2_{\rm ns} &=& \sum_{i=1}^{10} \left[{\avnsat(\lc[i])
- {\hat{N}_{\rm sat}}(\lc[i]) \over \Delta {\hat{N}_{\rm
sat}}(\lc[i])}
\right]^2 \,.\\
\end{eqnarray}
Here $\hat{X}$  denotes the  observational constraint $X$  and $\Delta
\hat{X}$ its  corresponding error. The  trial step is accepted  with a
probability given by
\begin{equation}
P_{\rm accept}=\left\{ \begin{array}{cl}
        1.0, & {\rm if} \chi^2_{\rm try} \le \chi^2_{\rm cur} \\
        {\rm exp}[-(\chi^2_{\rm try}-\chi^2_{\rm cur})/2], & {\rm if}
\chi^2_{\rm try} > \chi^2_{\rm cur}
           \end{array}\right.
\end{equation}
where $\chi^2_{\rm cur}$ denotes the $\chi^2$ for the current model in
the chain.

We initialize the chain from  a random position in the parameter space
and discard the first $10^4$  models allowing the chain to sample from
a more probable  part of the distribution. This  is called the burn-in
period  for the  chain. We  proceed and  construct a  chain  of models
consisting of  10 million models.  We thin this  chain by a  factor of
$10^4$ to  remove the  correlations between neighbouring  models. This
leaves  us with a  chain of  1000 independent  models that  sample the
posterior distribution.  We use this  chain of models to  estimate the
confidence levels on the parameters and relations of interest.

In Table~\ref{tab:table3},  we compare the $16^{\rm  th}, 50^{\rm th}$
and the $84^{\rm th}$  percentiles of the distributions of parameters,
which characterize  $\plcm$, obtained from  the MCMC with
the corresponding  true values of  these parameters present in  MOCKV. 
The  true  parameter  values  have  been  recovered  within  the  68\%
confidence intervals. The 68 and 95\% confidence levels in panels (a),
(b) and  (c) of Fig.~\ref{fig:fig6} show that  the models from
the  MCMC  accurately  fit  the  velocity  dispersions,  $\sigsw$  and
$\sighw$, as  well as  the average number  of satellites  per central,
$\avnsat$ as a function  of central galaxy luminosity.  The confidence
levels for the average luminosity of the centrals as a function of the
halo mass,  $\tilde{\lc}(M)$, are shown in panel  (d).  The confidence
levels on  the mean, $\avglogm$,  and the scatter, $\siglogm$,  of the
MLR of central galaxies are shown  in panels (e) and (f) respectively. 
The  solid lines  in  the  lower panels  show  the corresponding  true
relations present in MOCKV. Clearly,  our method is able to accurately
recover the true MLR.

This completes our tests with the MGC. Employing a variety of tests on
a realistic MGC, we have established a proof-of-concept that, starting
from a  redshift survey of  galaxies, one can reliably  select central
and  satellite  galaxies,  quantify  the kinematics  of  the  selected
satellites around  central galaxies and use this  information to infer
an unbiased estimate of the mean and the scatter of the MLR of central
galaxies.

\section{Results from Sloan digital sky survey analysis}
\label{sec:sdssapp}

We now apply  the method tested in the previous  sections to data from
the  Sloan Digital  Sky  Survey \citep[SDSS;][]{York2000}  which is  a
joint  five-passband  ($u,  g,  r,  i$ and  $z$)  imaging  and  medium
resolution ($R \sim 1800$) spectroscopic survey. More specifically, we
use   the   New  York   University   Value   Added  Galaxy   Catalogue
\citep{Blanton2005},  which   is  based  upon  SDSS   Data  Release  4
\citep{Adelman-McCarthy2006}  but   includes  a  set   of  significant
improvements over the original pipelines. From this catalog, we select
all galaxies  in the  main galaxy sample  with redshifts in  the range
$0.02 \le z  \le 0.072$ and with a  redshift completeness limit ${\cal
  C} >  0.8$. We  construct a volume  limited sample of  galaxies that
have a $r$-band luminosity (k-corrected to redshift 0.1) above $L_{\rm
  min}= 10^{9.5} \Lsunhh$. This sample is henceforth denoted by SDSSV.
It consists of  $57,593$ galaxies. The SDSSV data  is analysed assuming
the  cosmological parameters  from the  3  year data  release of  WMAP
\citep{Spergel2007},  $\Omega_{\rm  m}=0.238$, $\Omega_\Lambda=0.762$,
$h=H_0/100  \kms  \Mpc^{-1} =0.734$,  the  spectral  index of  initial
density    fluctuations    $n_{\rm    s}=0.951$   and    normalization
$\sigma_8=0.744$.

The iterative criterion  (ITER) outlined in Section~\ref{sec:selncrit}
is applied  to select  centrals and their  satellites from  SDSSV. The
parameters (a, b, c) in Eq.~(\ref{eq:sig200}) that define the aperture
used  in the final  iteration of  the central-satellite  selection are
listed in the last row  of Table~\ref{tab:table2}. The total number of
central  galaxies that  host at  least  one satellite  is $3863$.  The
number of satellite galaxies selected is $6101$.
\begin{figure}
\centerline{\psfig{figure=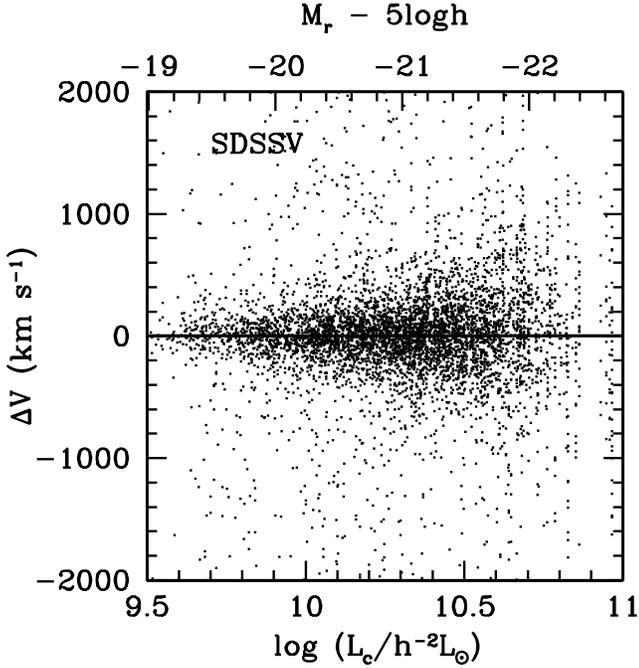,width=\hssize}}
\caption{Scatter plot of the velocity difference, $\dv$, between
  the  satellites and  their central  galaxies  as a  function of  the
  central galaxy  luminosity. The central galaxies  and satellites are
  selected  from   SDSSV  using  the   iterative  selection  criterion
  described in Section~\ref{sec:selncrit}.}
\label{fig:fig7}
\end{figure}
\begin{figure}
\centerline{\psfig{figure=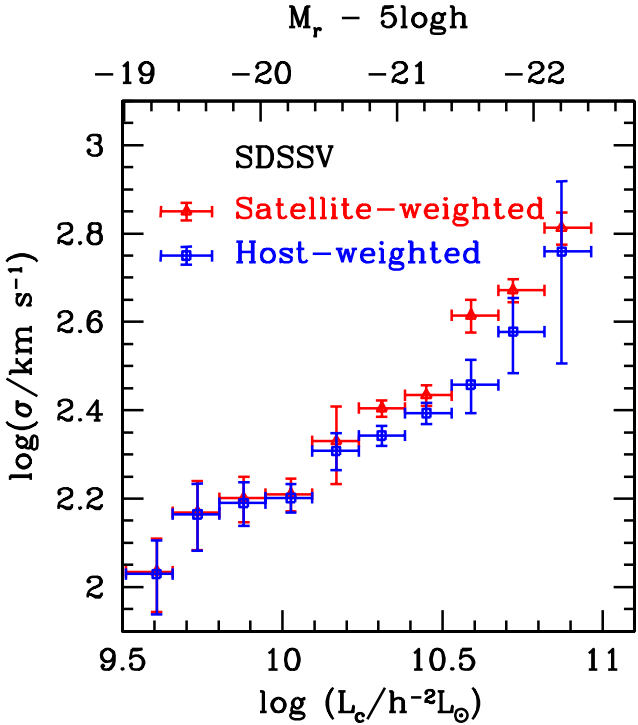,width=\hssize}}
\caption{ The satellite-weighted (red triangles) and the host-weighted
  (blue  squares)  velocity  dispersions  obtained from  centrals  and
  satellites selected from SDSSV.}
\label{fig:fig8}
\end{figure}
\begin{table}
\caption{SDSSV: Velocity Dispersion Measurements}
\begin{center}
 \begin{tabular}{rcccc}
   \hline
$\log(\lc)$ & $\sigsw$ & $\Delta \sigsw$ & $\sighw$ &
$\Delta \sighw$
\\
\hline
$\Lsunh$  & $\kms$ & $\kms$ & $\kms$ & $\kms$ \\
\hline\hline
 9.61   & 108 & 20 & 107 & 20 \\
 9.73   & 148 & 26 & 146 & 25 \\
 9.88   & 159 & 19 & 155 & 18 \\
10.03  & 162 & 14 & 159 & 12 \\
10.17  & 214 & 43 & 203 & 20 \\
10.31  & 254 & 11 & 220 & 11 \\
10.45  & 272 & 15 & 247 & 13 \\
10.59  & 412 & 35 & 287 & 39 \\
10.72  & 470 & 28 & 378 & 73 \\
10.87  & 650 & 54 & 574 & 254\\
\hline
\end{tabular}
\end{center}
\medskip

\begin{minipage}{\hssize}
  The velocity  dispersion measurements in  the satellite-weighted and
  host-weighted schemes together with the associated errors.
\end{minipage}
\label{tab:table4}
\end{table}

Fig.~\ref{fig:fig7}  shows  the  scatter  plot  of  the  velocity
difference,   $\dv$,   between  the   selected   satellites  and   the
corresponding centrals  as a function  of the central  luminosity. The
scatter in the velocities of satellites with respect to their centrals
clearly  increases with  central galaxy  luminosity. To  quantify this
scatter,   we   obtain   the   $\pdv$  distributions   in   both   the
satellite-weighting  and the host-weighting  schemes by  combining the
velocity differences,  $\dv$, of satellites within  luminosity bins of
uniform width  $\Delta \log [\lc] = 0.15$.  The satellite-weighted and
the  host-weighted  velocity  dispersions  are  estimated  from  these
distributions  by fitting  a double  Gaussian plus  constant  model as
described  in Section~\ref{sec:veldisp_b}. Fig.~\ref{fig:fig8}
shows  these dispersions  as  a function  of  central luminosity.  The
values of $\sigsw$, $\sighw$ and their associated errors are listed in
Table~\ref{tab:table4}.   Both    the   satellite-weighted   and   the
host-weighted velocity dispersions increase with the luminosity of the
central galaxy. Note  that the satellite-weighted velocity dispersions
are systematically higher than the host-weighted velocity dispersions.
As is  evident from Eqs.~(\ref{sweqnvol})  and ~(\ref{hweqnvol}), this
is a sufficient  condition to indicate the presence  of scatter in the
MLR of central galaxies (see Paper I for a detailed discussion).

\begin{figure*}
\centerline{\psfig{figure=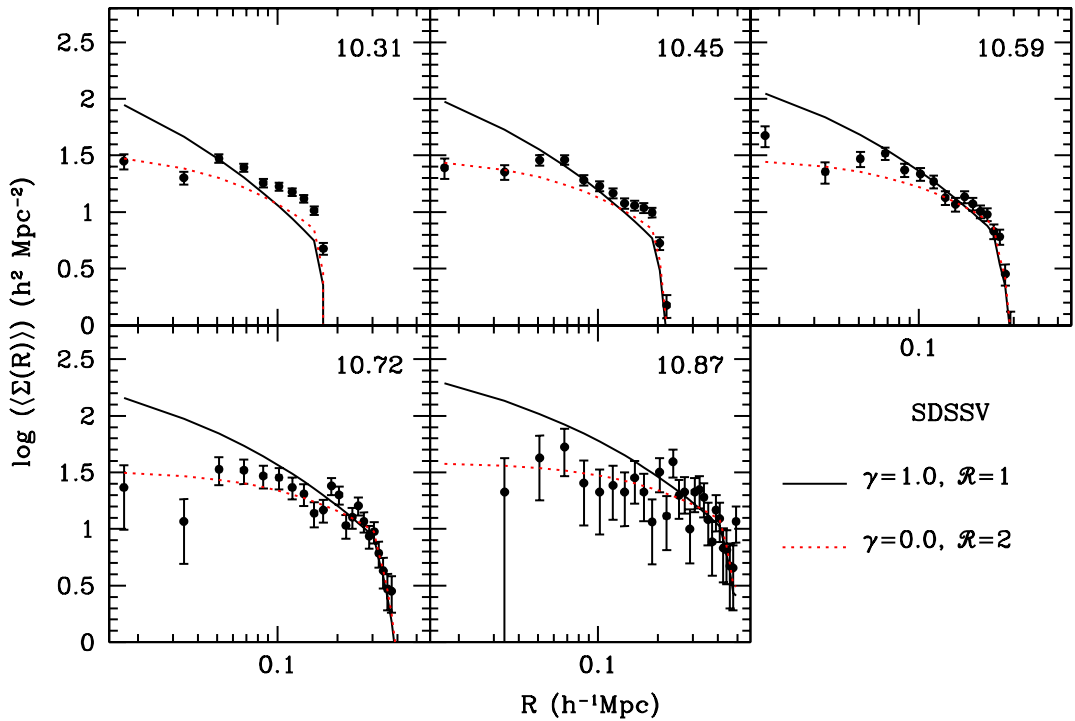,width=0.8 \hdsize}}
\caption{The projected number density distributions of
  satellites around  centrals selected from SDSSV for  the five bright
  luminosity bins  (average $\log (\lc/\Lsunh)$ in  right corner). The
  (black)  solid curves  indicate  the expected  distributions if  the
  number density  distribution of  satellites follows the  dark matter
  density,  i.e.  ${\cal R}=\gamma=1$  in  Eq.~(\ref{nsr}). The  (red)
  dotted  curves in  turn indicate  the expected  distributions  for a
  model  in  which  the  satellite   galaxies  are  a  factor  2  less
  concentrated than dark matter and  have a central core in the number
  density distribution i.e. ${\cal R}=2$, and $\gamma=0$.}
\label{fig:fig9}
\end{figure*}

The model  to infer  the MLR of  central galaxies from  the kinematics
requires the radial number density distribution of satellites, $n_{\rm
  sat}(r)$, as  an input.  For  inferring the MLR from  the kinematics
measured from MOCKV, we used  a model of $n_{\rm sat}(r)$ that follows
the density  distribution of  the dark matter  in an  unbiased manner,
i.e., $\gamma={\cal  R}=1$ in Eq.~(\ref{nsr}).  However  with SDSS, it
is not clear what functional form  of $n_{\rm sat}(r)$ should be used. 
In fact,  various studies have  shown that the satellite  galaxies are
spatially    antibiased   with    respect   to    the    dark   matter
\citep{Yang2005b,Chen2007a,Chen2007b}.  Rather  than 
including $\gamma$
and ${\cal R}$  as free parameters in our model,  we seek to constrain
these    parameters   using    the    observable   $\Sigma(R|\lc)$.    
Fig.~\ref{fig:fig9}   shows    the   projected   number   density
distributions  of  the  selected  satellites for  the  five  brightest
luminosity bins. As can be seen from Eq.~(\ref{satproflc}), predicting
$\Sigma(R|\lc)$  requires  the  knowledge  of $\pmlc$,  which  is  the
principle goal of our study. It further also requires the knowledge of
$\avnsatm$. Both these quantities are  unknown. We proceed as follows. 
We   use  the   $\pmlc$  and   $\avnsatm$  from   the  CLF   model  of
\citet{Cacciato2008}  which  was  also   used  to  populate  the  mock
catalogue. We  explore two different models for  $n_{\rm sat}(r)$, one
with  ${\cal R}=\gamma=1$,  where the  number density  distribution of
satellites follows the dark matter density distribution, and the other
with   ${\cal  R}=2$   and  $\gamma=0$,   where  the   number  density
distribution of satellites is spatially antibiased with respect to the
dark matter distribution. The former model is shown as a (black) solid
line   while    the   latter   with    a   (red)   dotted    line   in
Fig.~\ref{fig:fig9}.  Clearly, the  latter model is favoured by
the data.   Therefore, we use  ${\cal R}=2$ and $\gamma=0$  to specify
$n_{\rm sat}(r)$ for the analysis of the velocity dispersions to infer
the MLR of central galaxies.

\begin{figure*}
\centerline{\psfig{figure=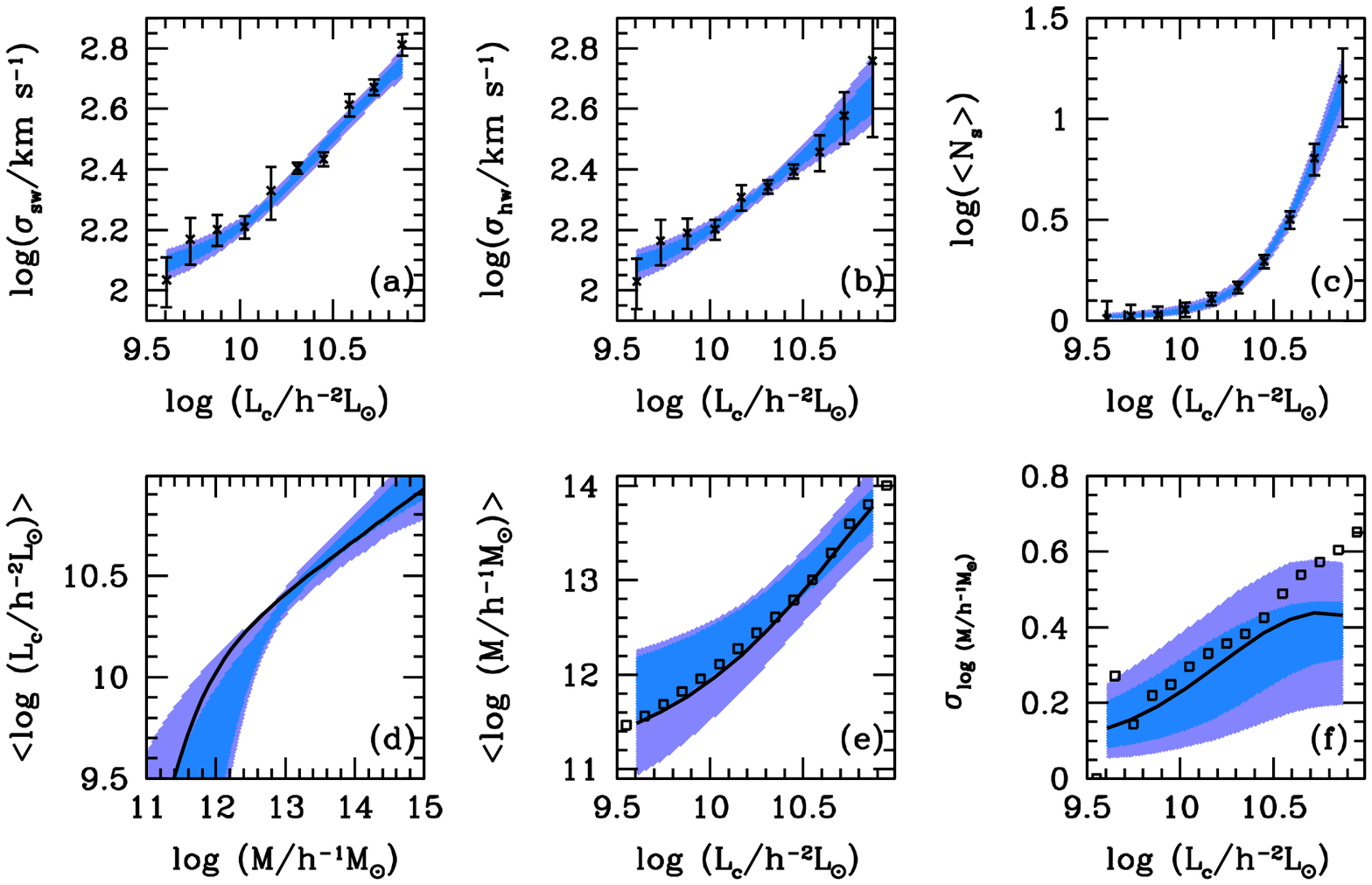,width= \hdsize}}
\caption{Crosses with errorbars in the upper panels show the data
  points used  to constrain the MCMC;  the satellite-weighted velocity
  dispersions in panel (a),  the host-weighted velocity dispersions in
  panel  (b)  and the  mean  number of  satellites  per  central as  a
  function  of luminosity  in panel  (c),  all measured  by using  the
  satellite  sample selected  from SDSSV.  The blue  and  purple bands
  represent  the 68\%  and 95\%  confidence regions  respectively. The
  bottom panels show the relations inferred from the MCMC; the average
  $\log(\lc)$ is in panel (d), and the mean and the scatter in the MLR
  of  central  galaxies  in  panels  (e)  and  (f)  respectively.  The
  relations obtained  by using  the halo occupation  distribution from
  MOCKV are shown using solid lines. The squares in panels (e) and (f)
  indicate  the  values  obtained  from the  semianalytical  model  of
  \citet{Croton2006}.  }
\label{fig:fig10}
\end{figure*}

Next     we    use     the    parametric     model     described    in
Section~\ref{sec:model_a} and constrain it using the measured velocity
dispersions, $\sigsw$  and $\sighw$, and the number  of satellites per
central, $\avnsat$ as  a function of the luminosity  of centrals. This
allows us to determine both the mean and the scatter of the MLR. Using
a  MCMC, we  recover  the mean  relations  $\avglogm$, $\siglogm$  and
$\avgloglc$.  The $16^{\rm  th}, 50^{\rm  th}$ and  the  $84^{\rm th}$
percentiles of the distributions of the parameter values that describe
the $\pmlc$ distribution for the central galaxies  are listed in
Table~\ref{tab:table5}. Fig.~\ref{fig:fig10} is the equivalent
of Fig.~\ref{fig:fig6}  but for  SDSSV. The upper  panels show
the data used  to constrain the parameters and  the bottom panels show
the confidence levels on the inferred mean and scatter of the MLR. The
values  of $\avglogm$  and $\siglogm$  together with  their 1-$\sigma$
errors are listed in Table~\ref{tab:table6}.

Clearly the average masses of dark matter haloes increase with central
galaxy luminosity,  as expected.   Interestingly, the scatter  in halo
masses  also  increases  systematically  with the  luminosity  of  the
central galaxy. At the bright end, this scatter is roughly half a dex.
Therefore, stacking central galaxies by luminosity amounts to stacking
haloes that cover a wide range  in masses.  This justifies the need to
account for this scatter in  the analysis of the satellite kinematics. 
Neglecting this scatter leads to an overestimate of the halo mass at a
given  central   luminosity.   Most  previous   studies  dealing  with
satellite kinematics have neglected this scatter which has resulted in
a  biased estimate of  the halo  mass-luminosity relationship.   As we
show  in  Appendix~\ref{sec:sample},  their  use of  strict  selection
criteria to  identify the centrals and satellites  have further biased
their estimate of the MLR of central galaxies.

\subsection{Comparisons with Independent Studies}

In  a recent  study, \citet{Cacciato2008}  have constrained  the   CLF
using the  abundance and  clustering of galaxies  in SDSS.   They have
shown  that this  CLF  is  also able  to  reproduce the  galaxy-galaxy
lensing  signal and is further consistent with the MLR obtained from a
SDSS group catalog \citep{Yang2008}. It is interesting to compare
the results of their
study with  the results obtained  here from satellite  kinematics. The
solid lines in the  bottom panels of Fig.~\ref{fig:fig10} show
the mean  and the scatter of  the MLR obtained using  the best-fit CLF
model  from  \citet{Cacciato2008}.  Clearly,  these  results are  in
excellent  agreement   with  the  results  obtained   here  using  the
kinematics of satellite galaxies. Amongst others this provides
further support that the halo mass assignment in the SDSS group
catalog of \citet{Yang2007} is reliable \citep{Wang2007b}.

Since the  mean and  scatter of  the MLR reflect  the physics,  and in
particular the  stochasticity, of galaxy formation,  it is interesting
to   compare   the  results   obtained   here   to  predictions   from
semi-analytical models  (SAM) of galaxy  formation. To that  extent we
use the SAM  of \citet{Croton2006}, which has been  shown to match the
observed  properties of  the local  galaxy population  with reasonable
accuracy\footnote{Note  that  \citet{Croton2006}  adopted  a  slightly
  different cosmology than the one used in our data analysis which can
  have a small  impact on the MLR.}. Using a  volume limited sample of
galaxies selected from  the SAM with the same  luminosity and redshift
cuts  as  SDSSV,   we  measure  the  mean  and   the  scatter  of  the
distributions of halo  masses for central galaxies in  several bins of
$r$-band luminosity.  The  results are shown in panels  (e) and (f) of
Fig.~\ref{fig:fig10} as open squares. The agreement with our
constraints from  the satellite kinematics is remarkably  good.  It is
both interesting  and encouraging that a  semi-analytical model, which
uses  simple, physically  motivated recipes  to model  the complicated
baryonic  physics  associated  with   galaxy  formation,  is  able  to
reproduce not  only the mean of  the MLR of central  galaxies but also
the correct amount of stochasticity in this relation.
\begin{figure}
\centerline{\psfig{figure=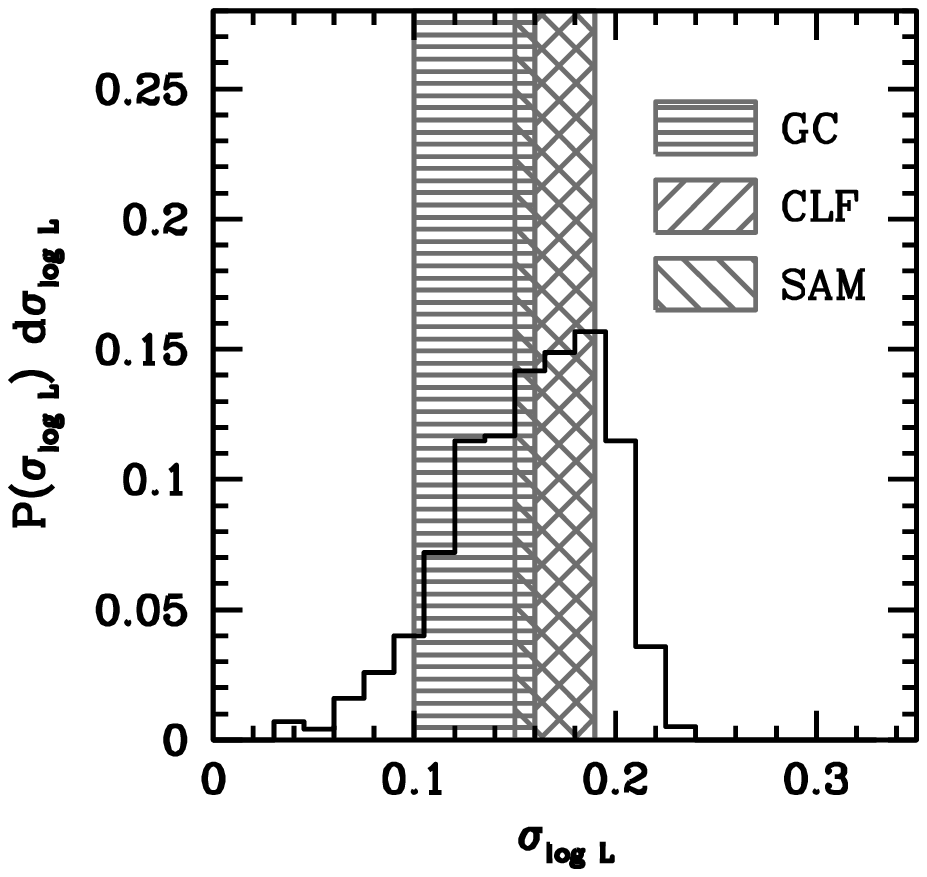,width=\hssize}}
\caption{The posterior distribution of the parameter $\sigcen$
  as  obtained  from  the  MCMC  analysis of  the  satellite  velocity
  dispersions. The 1-$\sigma$ constraints on the parameter $\sigcen$
  obtained from other independent methods are shown as shaded
  regions. Region GC indicates the SDSS group  catalog result by
  \citet{Yang2008}, region CLF indicates the result obtained  by 
  \citet{Cooray2006}  with  an independent CLF analysis and region
  SAM shows our measurement of $\sigcen$  from the  semi-analytical
  model of \citet{Croton2006}.}
\label{fig:fig11}
\end{figure}

In our model, the stochasticity  of galaxy formation is best described
by the parameter  $\sigcen$, which indicates the amount  of scatter in
the luminosity of central galaxies given  the mass of a halo, i.e. the
scatter   in    the   distribution   $\plcm$.     The   histogram   in
Fig.~\ref{fig:fig11} shows the  posterior probability of $\sigcen$,
obtained from our Monte Carlo Markov Chain, which yields that $\sigcen
= 0.16 \pm 0.04$ (68\% confidence levels).  Note that we have made the
assumption  that  $\sigcen$ is  independent  of  halo  mass. The  same
assumption was made by  \citet{Cooray2006}, who obtained that $\sigcen
= 0.17^{+0.02}_{-0.01}$  using the luminosity  function and clustering
properties of SDSS  galaxies \citep[see also][]{Cacciato2008}. Using a
large  SDSS galaxy group  catalogue, \citet{Yang2008}  obtained direct
estimates of the scatter in $\plcm$, and found that $\sigcen =0.13 \pm
0.03$  with no  obvious dependence  on  halo mass.   Finally, we  also
determined $\sigcen$  in the SAM of  \citet{Croton2006}: using several
bins in halo mass covering the range $10^{10} \Msunh \le M \le 10^{16}
\Msunh$,  we find  that $\sigcen  = 0.17  \pm 0.02$,  once  again with
virtually no dependence on halo mass. All these results are summarized
in Fig.~\ref{fig:fig11}.  Not only  do they support  our assumption
that  $\sigcen$  is  independent  of  halo  mass,  they  also  are  in
remarkable quantitative agreement with each other.  

Finally  we   note  that  a   constant  scatter,  $\sigcen$,   in  the
distribution  $\plcm$   leads  to   a  scatter,  $\siglogm$,   in  the
distribution $\pmlc$ that increases systematically with the luminosity
of the  central galaxy.  This arises  from the fact  that the relation
$\avgloglc$ (Panel  (d) in Fig.~\ref{fig:fig10})  is shallower
at the high mass end compared to the low mass end (see Paper I).

\begin{table}
\caption{SDSSV: Parameters recovered from the MCMC}
\begin{center}
 \begin{tabular}{cccc}
   \hline
Parameter & 16\% & 50\% & 84\%
\\
\hline\hline
$\log (L_0)$ &  9.69  &  10.05  &  10.33 \\
$\log (M_1)$ & 10.85  &  11.74  &  12.01 \\
$\gamma_2$   &  0.19  &   0.28  &   0.35 \\
$\sigcen$    &  0.12  &   0.16  &   0.19 \\
\hline
\end{tabular}
\end{center}
\medskip
\begin{minipage}{\hssize}
  The $16^{\rm  th}$, $50^{\rm th}$ and the  $84^{\rm th}$ percentiles
  of  the  distributions  of  parameters that  describe  the  relation
  $\tilde{\lc}(M)$  obtained from  the MCMC  analysis of  the velocity
  dispersions obtained from SDSSV.
\end{minipage}
\label{tab:table5}
\end{table}
\begin{table}
\caption{SDSSV: MLR of Central Galaxies}
\begin{center}
 \begin{tabular}{rcccc}
   \hline
$\log(\lc)$ & $\avg{\rm \log M}$ & $\Delta \avg{\rm \log M}$ &
$\sigma_{\rm \log M}$ &
$\Delta \sigma_{\rm \log M}$
\\
\hline
$\Lsunh$  & $\Msunh$ & $\Msunh$ & $\Msunh$ & $\Msunh$ \\
\hline\hline
   9.61  &  12.06  &   0.35  &   0.12  &   0.06 \\
   9.73  &  12.16  &   0.32  &   0.13  &   0.07 \\
   9.88  &  12.28  &   0.29  &   0.15  &   0.08 \\
  10.03  &  12.44  &   0.26  &   0.18  &   0.10 \\
  10.17  &  12.60  &   0.23  &   0.22  &   0.10 \\
  10.31  &  12.80  &   0.21  &   0.26  &   0.11 \\
  10.45  &  13.01  &   0.19  &   0.30  &   0.10 \\
  10.59  &  13.24  &   0.19  &   0.34  &   0.09 \\
  10.72  &  13.47  &   0.21  &   0.36  &   0.08 \\
  10.87  &  13.74  &   0.23  &   0.38  &   0.07 \\
\hline
\end{tabular}
\end{center}
\medskip

\begin{minipage}{\hssize}
  The mean and scatter of the halo masses as a function of the central
  galaxy  luminosity inferred from  the MCMC  analysis. The  errors on
  each of  the inferred quantities  correspond to the  68\% confidence
  levels.
\end{minipage}
\label{tab:table6}
\end{table}

\section{Summary}
\label{sec:summary}

The  kinematics  of  satellite  galaxies  have  been  widely  used  to
statistically relate the mean halo masses of central galaxies to their
luminosities (Zaritsky \etal 1993; 
Zaritsky \& White 1994; Zaritsky \etal 1997; McKay \etal 2002; Prada
\etal 2003; Brainerd \& Specian 2003; Norberg \etal 2008). These
studies use strict criteria to
identify central and satellite  galaxies that reside preferentially in
isolated environments. Following \citet{van den Bosch2004}, we applied
a relaxed but adaptive selection  criterion to a volume limited sample
from  SDSS to  identify centrals  and  their satellites,  not only  in
isolated environments  but also in  massive groups and  clusters which
allowed us to study the kinematics  of satellites over a wide range of
central galaxy luminosity.  We inferred both the mean  and the scatter
of  the  mass-luminosity relationship  of  central  galaxies from  the
kinematics of satellite galaxies.

Using a  realistic mock catalogue,  we thoroughly tested  the analysis
method at  every step. We tested  the performance of  our criterion to
identify central and satellite galaxies  and our method to measure the
kinematics of  the selected satellites. We showed  that the kinematics
recovered from the selected satellites are a fair representation of
the
true kinematics of  satellite galaxies present in the  mock catalogue. 
We  presented  an analytical  model  that  properly  accounts for  the
selection biases  and showed that  the predictions of  this analytical
model  are in  good  agreement  with the  measured  kinematics of  the
selected satellites.

In Paper I,  we have shown that the  velocity dispersion of satellites
can   be    measured   using   two    different   weighting   schemes:
satellite-weighting  and  host-weighting.    We  have  demonstrated  a
degeneracy between the  mean and the scatter of  the MLR obtained from
either the satellite-weighted or the host-weighted velocity dispersion
alone. However, we have also  shown that this degeneracy can be broken
by using the  velocity dispersions in the two  schemes simultaneously. 
In  this  paper  we first  tested  our  method  using a  mock  galaxy
catalogue. We fitted the measured satellite-weighted and host-weighted
velocity dispersions  simultaneously using a parametric  model for the
halo  occupation statistics  of  central and  satellite galaxies,  and
demonstrated that we can reliably obtain confidence levels on the true
mean and scatter of the mass-luminosity relation of central galaxies.

Next we applied  the above method to a volume  limited sample from the
SDSS. The mean of the mass-luminosity relation increases as a function
of  the central  host luminosity  indicating that,  as  expected, more
massive haloes host brighter centrals. This result is in excellent
quantitative agreement with a recent study by \citet{Cacciato2008},
who use the abundance and clustering properties of galaxies in SDSS to
constrain the CLF, and with the SAM of \citet{Croton2006}. The
satellite kinematics obtained in our study are
consistent with a model in which $\plcm$ has a constant scatter,
$\sigcen$, independent of the halo mass $M$. We obtain
$\sigcen=0.16\pm0.04$ in excellent agreement with other independent
measurements suggesting that the amount of stochasticity in galaxy
formation is similar in haloes of all masses. This is also suggested
by the SDSS group catalog of \citet{Yang2008} and by the SAM of
\citet{Croton2006}.

\section*{Acknowledgments}

We are grateful to Kris Blindert, Jacqueline Chen, Anupreeta More,
Peder Norberg, Hans-Walter Rix and Joachim Wambsganss for valuable
discussion.

\appendix
\section{Sampling of central galaxies}
\label{sec:sample}

\begin{figure*}
\centerline{\psfig{figure=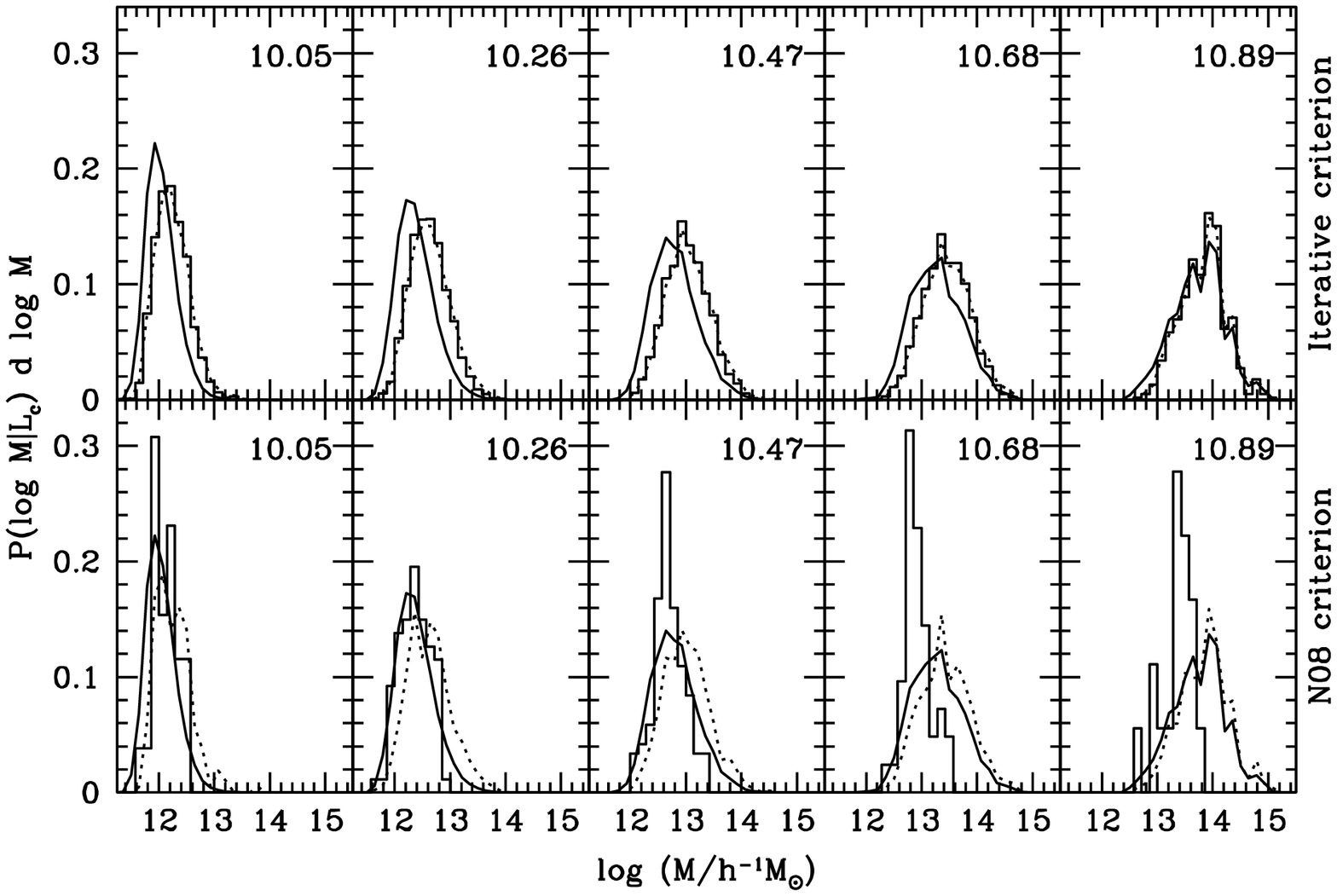,width=0.8\hdsize}}
\caption{Comparison of the sampling of central galaxies using the
  iterative  selection criterion  (ITER) used  in this  paper  and the
  criterion used in Norberg \etal  (2008) (N08). The histograms in the
  upper  (bottom)  panel show  the  distributions  of  halo masses  of
  central galaxies selected according to ITER (N08). The average $\log
  (\lc / \Lsunh)$ for central galaxies in each bin is indicated at the
  top  right corner  of  each panel.  The  solid lines  show the  true
  distributions of halo masses for all the central  galaxies and the 
  dotted lines show the distribution of halo masses of those central 
  galaxies that have  at least  one satellite  more than  $f_{\rm s}$
  times fainter than  themselves in the inner cylinder  defined by the
  selection criterion.}
\label{fig:fig12}
\end{figure*}

The  ultimate  goal of  satellite  kinematics  is  to probe  the  halo
mass-luminosity relationship (MLR)  of central galaxies. In principle,
an unbiased  estimate for the  MLR requires that the  central galaxies
identified  by the  selection criterion  are an  unbiased (sub-)sample
with  respect  to their  corresponding  dark  matter  haloes. In  this
appendix, we  investigate, using the MGC, how  our iterative criterion
performs in this respect and compare it with the strict criterion used 
in the literature.

For  reasons that  will become  clear later,  we use  the flux-limited
sample    MOCKF    for    this    test.    The    solid    lines    in
Fig.~\ref{fig:fig12}  show  the  distributions  of  halo  masses,
$\pmlc$, for  all central  galaxies in MOCKF  divided in  5 luminosity
bins. The average logarithm of the luminosities of  central galaxies
in
each bin is  indicated at the top right corner.  The dotted lines show
the  distributions, $\pmlc$,  for all  central galaxies  that  have at
least one satellite in the selection aperture defined by our iterative
selection criterion.  Finally, the histograms  show the distributions,
$\pmlc$, of the centrals selected by our iterative criterion. Clearly,
the  centrals   selected  by   our  iterative  criterion   sample  the
distribution of halo  masses from the dotted lines  (and not the solid
lines).  However, as  discussed  in Section~\ref{sec:veldisp_c},  this
bias  is  taken  into  account  while modelling  the  kinematics  (see
Eq.~[\ref{poisson}])  and  therefore allows  us  to  make an  unbiased
estimate. As  shown in  Section~\ref{sec:model}, we indeed  recover an
unbiased MLR  from the kinematics of the  selected satellites measured
around the centrals selected by our iterative criterion.

For comparison, we now repeat this exercise using the strict criterion
employed in  previous studies. In  particular, we adopt  the criterion
used in N08. This criterion identifies  a galaxy as a central if it is
at least $f_{\rm h}=2$ times brighter than any other galaxy in a fixed
(irrespective of the luminosity  of the galaxy) aperture cylinder (see
Table~\ref{tab:table1}) around  itself.  Satellites are  identified as
those galaxies that are at  least $f_{\rm s}=8$ times fainter than the
centrals and reside in a  smaller aperture cylinder defined around the
centrals.  The  values of  $f_{\rm  h}$ and  $f_{\rm  s}$  in the  N08
criterion are conservative,  as the principle goal of  their study was
to  select isolated central  galaxies. Applying  the N08  criterion to
MOCKV  selects only  $126$ satellites  around $96$  central  galaxies. 
Therefore, to do  a meaningful comparison, we apply  the N08 criterion
to MOCKF for which it  selects $657$ satellites around $395$ centrals. 
For  comparison,  our iterative  criterion  yields $39951$  satellites
around $21206$ centrals.

Solid lines  in the lower panels of  Fig.~\ref{fig:fig12} are the
same as in the upper panels and show the distributions of halo masses,
$\pmlc$,  for all  central  galaxies  in MOCKF  divided  in 5  central
luminosity bins. The dotted lines  show the $\pmlc$ for those centrals
that have at least one  satellite around them which is $f_{\rm s}(=8)$
times fainter than themselves.  There is a negligibly small difference
in the dotted lines in the two rows due to different values of $f_{\rm
  s}$. Finally,  the histograms show the $\pmlc$  distributions of the
sample of centrals  selected by the N08 criterion.   Clearly, these do
not  sample  the distributions  shown  by  the  dotted lines  and  the
distributions are clearly biased towards the low mass end, especially,
in the  bright luminosity  bins.  This owes  to the fact  that Norberg
\etal  (2008) adopt  $f_{\rm  h} =  2$,  which preferentially  selects
centrals that do not have satellite galaxies of comparable brightness.
This biases the distributions towards the low mass end.  Note, though,
that this is not a  critique regarding their selection criteria; after
all, as Norberg  \etal (2008) clearly described in  their paper, their
principal  goal  is to  study  the  kinematics  around {\it  isolated}
galaxies. However, it  does mean that it is  not meaningful to compare
their  MLR, which  is only  applicable to  isolated galaxies,  to that
obtained here,  which is representative  of the entire  central galaxy
population.

\end{document}